\documentclass[10pt, pre, aps, twocolumn, superscriptaddress]{revtex4-1}
\pdfoutput=1
\usepackage{amsmath,natbib,amsfonts}
\usepackage{amssymb,color, float}
\usepackage{graphicx}
\usepackage{caption}
\usepackage{ragged2e} 
\makeatletter
\long\def\@makecaption#1#2{%
  \vskip\abovecaptionskip
  \sbox\@tempboxa{#1. #2}%
  \ifdim \wd\@tempboxa >\hsize
    \justifying #1. #2\par
  \else
    \justifying \hbox to\hsize{\hfil\box\@tempboxa\hfil}\par
  \fi
  \vskip\belowcaptionskip}
\makeatother
\usepackage{subfig}
\usepackage{tikz}
\usepackage[colorlinks=true, citecolor=blue, urlcolor = blue, linkcolor= red,bookmarks=true]{hyperref}
\newcommand{\beqa}{\begin{eqnarray}}
\newcommand{\eeqa}{\end{eqnarray}}

\begin{document}
\title{Cyclic machines in contact with thermal and athermal baths: \\ the role of Poisson shot noise}
\author{Rita Majumdar} 
\affiliation{Department of Physics, Indian Institute of Technology Delhi, Hauz Khas, 110 016 New Delhi, India}
\affiliation{Biological Complexity Unit, Okinawa Institute of Science and Technology (OIST),
1919-1 Tancha, Onna, Okinawa 904-0495, Japan}
\affiliation{ICTP -- The Abdus Salam International Centre for Theoretical Physics, 34151 Trieste, Italy}
\author{Costantino Di Bello}
\affiliation{Scuola Internazionale Superiore di Studi Avanzati, 34136 Trieste, Italy}
\author{Ralf Metzler}
\affiliation{Institute of Physics \& Astronomy, University of Potsdam, 14476 Potsdam, Germany}
\author{Rahul Marathe}
\email{maratherahul@physics.iitd.ac.in}
\affiliation{Department of Physics, Indian Institute of Technology Delhi, Hauz Khas, 110 016 New Delhi, India}
\author{\'Edgar Rold\'an}
\email{ edgar@ictp.it}
\affiliation{ICTP -- The Abdus Salam International Centre for Theoretical Physics, 34151 Trieste, Italy}

\date{\today}
\begin{abstract}
We study stochastic models of a microscopic active heat engine, comprised of an overdamped Brownian particle trapped in a harmonic potential, and in simultaneous contact with thermal (passive)  and  athermal (active) baths. The interaction with the active bath is modeled as a stochastic force described by Poisson shot-noise (PSN) having a specified amplitude distribution. With analytical calculations and numerical simulations, we study the thermodynamic performance of this machine with cyclic protocols that are analogous to two-stroke and Stirling-like engines. For specific parameter ranges, the thermodynamic behavior is that of a hybrid machine, operating as a heat engine with respect to the thermal (passive) bath and as a refrigerator with respect to the PSN (active) bath, or vice versa. Focusing on the parameter region where the overall performance is that of an engine, we show that the average total extracted work per cycle divided by average total heat intake from the hot baths per cycle may surpass the Carnot efficiency associated with the temperature of the passive baths. Applying the second law for active heat engines, we focus on a bonafide efficiency  (bounded by Carnot's efficiency) that incorporates an information-theoretic metric~$\mathcal{I}$ ---which we call  {\em quasistatic divergence}--- quantifying how distinguishable are the engine's statistics in the quasistatic limit with respect to a continually changing equilibrium distribution. We analyze, with theory and numerical simulations, how the PSN shot rate and the degree of non-Gaussianity in the particle position distribution influence the efficiency of the engine, and explore the  correlation between non-Gaussianity and efficiency. Our findings reveal optimal PSN shot rates maximizing the engine's efficiency and an intriguing non-bijective relation between efficiency and kurtosis. 
  \end{abstract}
\maketitle  

\section{Introduction}
\label{S1}
Fundamental concepts and laws of thermodynamics form the basis for understanding key physical phenomena and engineering applications regarding the how machines such as engines and refrigerators work. 
For an idealized Carnot engine operating quasi-statically 
between two thermal baths at temperatures $T_c$ (cold bath) and $T_h$ (hot bath), the efficiency, defined by the ratio between the extracted work per cycle and the heat absorbed from the hot bath per cycle, equals to the so-called Carnot value  $\eta_C=1-(T_c/T_h)$~\cite{Carnot}.
Carnot's efficiency sets the fundamental upper limit for the efficiency attainable by any macroscopic heat engine, being attainable only through the combination of isothermal and adiabatic processes and in the reversible (quasistatic) limit, which in practice results in a net zero power output.

Miniaturized systems characterized by few nonequilibrium degrees of freedom, such as colloidal particles and molecular motors, often operate near the fundamental limits--- with energy fluxes with the same order of magnitude as the  thermal energy $k_B T$, with $k_B$ Boltzmann's constant and $T$ the temperature of the environment.  Hence, the usage of the modern framework of {\em stochastic} thermodynamics is mandatory to describe the fluctuations of work, heat, and entropy fluxes associated with the erratic (stochastic) motion of small machines.
Recent 
experiments with a variety of microscopic machines 
have revealed intriguing statistical properties associated with the thermodynamic quantities characterizing the performance of small machines~\cite{Ciliberto}. In particular, single-particle heat engines have become a modern paradigm over the last two decades through experiment-theory verifications of unprecedented accuracy in e.g. colloidal, atomic, and biomolecular systems~\cite{Martinez16,Robnagel16}.
Stochastic thermodynamics, fluctuation theorems 
provide generalized second laws for systems driven arbitrarily away from equilibrium. When applied to  mesoscopic heat engines undergoing deterministic cyclic protocols, the second law of stochastic thermodynamics for the entropy production imposes a bound on the efficiency  $\eta$ defined as in macroscopic thermodynamics, yet in terms of work and heat {\em averaged over} many cycles. In particular, $\eta=\langle w \rangle/\langle q_h \rangle \leq \eta_C$, where $\langle w \rangle$ and $\langle q_h \rangle$ denote respectively the  extracted work from the engine and  the heat intake by the engine from the hot bath, both averaged $\langle \;\cdot\;\rangle $ over many realizations of  the engine's cycle~\cite{Martinez16,Rana2016}.

The study of the so-called {\em active matter}  has gained significant attention  over the last few decades, examples include flocks of birds, interacting (natural or artificial) microswimmers, or biological systems among others ~\cite{Vicsek12, Ramaswamy10,Cates, Volpe16,Goswami23}. In short, active matter systems are more than just a material~\cite{khodabandehlou2026bringing}, as they are characterized by an autonomous, often steady expenditure of fuel energy and its conversion into efficient information-processing and work-production output tasks. The energy continuously consumed by active matter can also be harvested and converted into useful work, for example by using swimming bacteria to drive bacterial ratchets or powering microscopic heat engines with bacterial reservoirs~\cite{DiLeonardo2010,Krishnamurty16}.
Researchers have extended stochastic thermodynamics for active systems \cite{leticia,Maes14,Maes15,Cates,Genesotto18,Edgar,lahiri2024efficiency}, providing insights into their dynamics. Seminal experimental work~\cite{Krishnamurty16} reported the realization of a Stirling heat engine  with a colloidal microscopic particle immersed in a bacterial bath, suggesting enhancements of power and efficiency with respect to equilibrium thermodynamics' predictions. Theoretical work studied thoroughly `active' heat engines   \cite{Marathe18, Marathe19, Marathe22, Ruben23}, tackling their thermodynamics aspects  by mapping their fluctuations to  passive systems in a high (effective) temperature, defining a dynamic effective temperature, and the Carnot efficiency in terms of such effective temperatures \cite{Viktor20,Viktor_Rahul_20,Kroy24,Kroy241}. 
Such approaches motivate further developments, especially when considering non-Gaussian and/or time-correlated fluctuations that may not be explained using only Gaussian white noise, even at high effective temperatures. 
In this regard, a pioneering  theoretical framework (focused on the notion of excess entropy) has introduced  a bonafide efficiency $\eta_a$   suitable  for  active heat engines, and more generally for heat engines in contact with thermal baths and non-conservative forces. The  key aspect of the `active' efficiency  $\eta_a$ is that it is always bounded by the Carnot limit associated with the temperature of the thermal bath/s in which the source/s of activity are embedded~\cite{Barato}, see also~\cite{Baratoref26} for an extension to  refrigerators. As we will show later, an efficient evaluation of the active efficiency relies on quantifying an information-theoretic quantity $\mathcal{I}$, which we call {\em quasistatic divergence} (see  Eq.~\eqref{eq:I_inf}) that is reminiscent of the concept of lag~\cite{vaikuntanathan2009dissipation} in irreversible processes and of the so-called mismatch cost~\cite{manzano2024}. The quasistatic divergence~$\mathcal{I}$ quantifies the distance between the statistics of the engine under quasistatic driving and the continually evolving equilibrium statistics   that it would follow in absence of active noise.    While being a modern theory of (very) broad applicability, the efficiency of active heat engines has not been fully exploited yet despite the availability of a plethora of experimental data of passive probes immersed in active baths~\cite{Krishnamurty16, Wu20, Albay21, Krishnamurty21, Cheng22}.

Mainstream stochastic theory of active systems and passive systems in active baths often rely in minimal models such as run-and-tumble and active Ornstein Uhlenbeck processes, which describe persistent nonequilibrium motion in a simple way. In run-and-tumble models, a particle moves in nearly straight ``runs" and randomly changes direction through ``tumbles", while in active Ornstein-Uhlenbeck models the active force is represented by a persistent exponentially correlated noise~\cite{TailleurCates2008,CatesTailleur2015,SolonCatesTailleur2015,Bonilla2019,DabelowEichhorn2021,Martin2021}. 
A recent, yet unorthodox  approach to describe passive probes in contact with active-matter baths  is to consider Langevin dynamics with both Gaussian white (thermal) and Poisson shot noise (PSN)~\cite{Edgar24, Goswami21, GOELPSNescapePhysicaA2026}. The rationale behind PSN is to introduce a  force exerted  as pulses (`kicks') at  stochastic times and each with a randomly-distributed  amplitude. Such kicks may represent approaches and interactions of the probe with of active particles present in a dilute solution of bacteria~\cite{Edgar24}.  See also \cite{SpiechowiczPRE25, SpiechowiczChaos25, LiebchenJCP24, Sarkar2025, BanerjeePRE25, RoichmanSoft26}, for recent progress in non-equilibrium statistical physics  in the presence of PSN or active noises. 
In this paper, we follow up our recent work by putting forward  isothermal shot-noise  models~\cite{Edgar24} to consider a stochastic heat engine in the presence of both thermal (Gaussian white noise) and active (PSN) fluctuations.
We investigate the impact of  the PSN statistics  on the performance efficiency of a minimalistic model of an active heat engine. As a key milestone, we unveil the possibility of performance as 
"hybrid machines"~\cite{manzano2020hybrid}, operating as a heat engine with respect to the thermal bath and as a
refrigerator with respect to the PSN bath, or, in a different parameter
regime, as a refrigerator with respect to the thermal bath and as a heat
engine with respect to the PSN bath. By considering  cyclic processes comprised of  isothermal and adiabatic steps, we explore the second law derived from the excess entropy for active systems and its applicability to our model. With this model, we aim to test whether increasing the degree of non-Gaussianity in the system's statistics (excess kurtosis)  improves the engine's efficiency, as suggested by the experimental work with bacterial reservoirs~\cite{Krishnamurty16}. Similarly, our model aims to tackle experimental scenarios with synthetic (PSN) noise~\cite{Krishnamurty21} generated for example by optical tweezers or external fields, and  advance the understanding of the efficiency of active heat engines.
The rest of this work is organized as follows. In Sec.~\ref{sec:Quantifying_eficiency_in_active_engine}, we introduce our setup by describing the motion of a trapped Brownian particle subjected active bath modeled by the Poisson shot noise. In Sec. \ref{stocengergy}, we discuss stochastic energetics, employ two separate definitions and interpretations of engine efficiency, and examine the active efficiency, considering the heat dissipation caused by the active bath. In Sec.~\ref{sec:Two_stroke_engine} and Sec.~\ref{sec:Two_stroke_engine2}, we discuss a two-step protocol to operate as an engine. In this case, we  obtain analytical and numerical  results for key thermodynamic quantities and the engine's active efficiency by considering the Gaussian kick distributions. In Sec.~\ref{sec:Discussion_final}, we summarize our findings providing a comprehensive overview of our results and their implications.
The Appendices  contain  details of calculations used in the Main Text as well as additional results.

\section{System dynamics and  Fokker-Planck equation solution}
\label{sec:Quantifying_eficiency_in_active_engine}
We study the dynamics of an overdamped Brownian particle immersed in a thermal bath with time-dependent temperature that follows a deterministic protocol $T(t)$ within two values $T_c$ (cold) and $T_h>T_c$ (hot). The particle is subject to a time-dependent one-dimensional harmonic potential $U(x,t)=\kappa(t)x^2/2$ with $\kappa(t)$  the  stiffness at time $t$. 
In addition, the particle  interacts with an  active bath through a stochastic active force $\eta(t)$ whose statistical properties are described below. The position of  the  particle obeys the  overdamped Langevin equation
\begin{eqnarray} 
\label{eqn:Langevin_2}
\gamma\dot{x}(t)=-\kappa(t)x(t)+\sqrt{2k_BT(t)\gamma}~\xi(t)+\gamma\zeta_a(t).
\end{eqnarray}
Here, $\gamma$ is the friction coefficient; the thermal fluctuation is modeled as Gaussian white noise $\xi(t)$ with zero mean $\langle\xi(t)\rangle=0$  and autocorrelation $\langle\xi(t)\xi(t')\rangle=\delta(t-t')$. 

The active noise $\zeta_a(t)$ is modeled as a Poisson shot noise (PSN) characterized by the exertion of random forces (kicks) at random (Poissonian) times. Specifically, the random times $t_i$ elapsed between two successive kicks are independent and exponentially distributed with rate $\omega_a$ (see \cite{hanggi78, hanggi80} for a generalization of the model to PSN with finite duration pulses), while the kicks amplitudes $Y_i(t)$ are independent Gaussian random variables. The distribution of $Y_i(t)$, here denoted by $\rho_a(y,t)$, reads
\begin{equation}
    \rho_a(y,t) = \frac{\exp(-y^2/2\sigma^2_a(t))}{\sqrt{2\pi\sigma^2_a(t)}},
    \label{kickGauss}
\end{equation}
with $\sigma^2_a(t) = \langle Y_i^2 (t)\rangle\  $ the variance of the $i-$th kick amplitude distribution which may also be time-dependent. 
The expression for the active noise is given by, 
\begin{eqnarray}
   \zeta_a(t)=\sum_{i=1}^{N_t}Y_i(t)\delta(t-t_i).\label{eq:PSN}
\end{eqnarray}
The mean and autocorrelation functions associated with the PSN noise Eq.~\eqref{eq:PSN} are given by (see~\cite{hanggi78})
\begin{eqnarray}
 \langle\zeta_a(t)\rangle&=&0,\\~\langle\zeta_a(t)\zeta_a(t')\rangle&=&\omega_a\sigma^2_a(t)\delta(t-t').
\end{eqnarray}
For the case of
a time-dependent stiffness $\kappa(t)$ and time-independent temperature $T$,
 the Fokker-Planck equation for the probability density $P(x,t)$ associated with the position $x$ reads ~\cite{hanggi78}
\begin{equation}
\label{eqn:FokkerPlanck}
    \begin{aligned}
    \dfrac{\partial}{\partial t}P(x,t) &= \dfrac{\kappa(t)}{\gamma} \dfrac{\partial}{\partial x}\left[x P(x,t)\right]+\frac{k_BT}{\gamma}\dfrac{\partial^2}{\partial x^2}P(x,t)\\
    &+\omega_a \int_{-\infty}^{\infty}[P(x-y,t)-P(x,t)]\rho_a(y,t)dy,
    \end{aligned}
\end{equation}
with vanishing probability at $\pm \infty$ at all times $t$. We derive the explicit form of $P(x,t)$ considering a delta initial condition ${P(x,t_0) = \delta(x-x_0)}$ using the method of characteristics (see Appendix~\ref{app:FPE_sol})
\begin{eqnarray}
    \label{eqn:FokkerPlanck_solution}
    P(x,t)& = &\int_{-\infty}^{+\infty} \dfrac{dq}{2\pi} \exp\biggr[ -iq \left( x-\dfrac{x_0}{\varphi(t,t_0)}\right)\\ & -&\dfrac{k_BT}{\gamma} q^2 \dfrac{\int_{t_0}^t \varphi(t',t_0)^2 dt'}{\varphi(t,t_0)^2} + \dfrac{\omega_a \gamma}{\kappa(t)} J(q,t,t_0) \biggr],\nonumber
\end{eqnarray}
where,
\begin{eqnarray}
    \label{eqn:varphi_def}
    \varphi(t,t_0) &=& \exp \left(\int_{t_0}^t \dfrac{\kappa(s')}{\gamma} ds' \right),\\
    J(q,t,t_0) &=& \int_{t_0}^t \left[\hat \rho_a \left( q \dfrac{\varphi(t',t_0)}{\varphi(t,t_0)} , t' \right) - 1 \right] dt'
    \label{eqn:varphi_def2}
\end{eqnarray}
Here, 
\begin{equation}
\hspace{-0.2cm}
    \hat{\rho}_a(z,t) = \int_{-\infty}^{\infty} e^{izy} \rho_a(y,t)~dy
    =\exp\left(-\frac{1}{2} \sigma^2_a(t) z^2 \right),
\end{equation}
is the Fourier transform of the kick amplitude distribution $\rho_a(y)$. As a sanity check, we find  that because $x(t)$ is a L\'evy process, its density $P(x,t)$ must fulfill  the so-called L\'evy-Khintchine representation \cite{zolotarev}. Formulae~(\ref{eqn:FokkerPlanck_solution}-\ref{eqn:varphi_def2}) are the first main results of the paper. To the authors' knowledge, the analytical solution of the time-dependent Fokker-Planck equation~\eqref{eqn:FokkerPlanck} is not available in the literature. Even though its analytical expression is highly-nontrivial,  formula~\eqref{eqn:FokkerPlanck_solution} represents the building block of most of the thermodynamic quantities discussed later in the paper.   The general solution derived above is not restricted to the piecewise-constant protocols considered in the Main Text, but have a  broader scope, see Appendix~\ref{app:stirling_protocol} for  Stirling-type engines where the    stiffness is modulated smoothly with time.

For our forthcoming analyses it will be convenient to introduce the notation
\begin{equation}
    \Lambda(t)=\{\kappa(t),T(t),T_a(t)\},
\end{equation}
to denote the value of the deterministic protocol at time~$t$ of  the control parameters, namely stiffness $\kappa$,  temperature of the thermal bath $T$, and effective temperature of the PSN $T_a$. 
A central quantity for the analysis exposed in the paper will be the accompanying density  $P^{\rm ac}$ (see~\cite{hanggi82, chetrite_epl}) defined as the time-dependent probability density function  that at all times $t$ obeys
\begin{equation}
    \mathcal{L}^\dagger P^{\rm ac}(x|\Lambda(t))=0,
\end{equation}
where $\mathcal{L}^\dagger$ is the Fokker-Planck operator governing the forward evolution of the system. In other words, the Fokker-Planck equation~\eqref{eqn:FokkerPlanck} can be rewritten in terms of this operator as $\partial_t P=\mathcal{L}^\dagger P$. Notice that $\mathcal{L}^\dagger P^{\rm ac}=0$ does not imply that $\partial_t P^{\rm ac}=0$, thus $P^{\rm ac}$ is not the stationary distribution. On the other hand, $P^{\rm ac}$ is the stationary distribution that may be reached if all the time-dependent parameters are frozen at their instantaneous value $\Lambda(t)$~\cite{hanggi82}. For this reason we preferred the notation $P^{\rm ac} (x|\Lambda(t))$ often used for conditional probability. The accompanying density here reads
\begin{eqnarray}
    P^{\rm ac}(x|\Lambda(t)) &=&\int_{-\infty}^{+\infty} \dfrac{dq}{2\pi} \exp \left(-iqx -\dfrac{k_BT(t)}{2 \kappa(t)}  q^2 \right.\nonumber\\
    &+& \left. \dfrac{\omega_a \gamma}{\kappa(t)}\int_{0}^{q} \dfrac{\hat{\rho}(z,t)-1 }{z} dz \right). 
    \label{eqn:ps_noneqsteady} 
\end{eqnarray}  
This formula is already known in the literature in the case of constant parameters \cite{eliazar_klafter, Edgar24, kanazawaPRL} and it will useful later in the paper. 

The presence of the  PSN active noise implies that  the accompanying density is in general 
non-Gaussian and thus different from the equilibrium (Boltzmann) distribution associated with the protocol value $\Lambda(t)$.  In particular, for the zero-mean Gaussian kick distribution given by Eq.~\eqref{kickGauss}, $P^{\rm ac}$ has zero mean $ \langle x\vert \Lambda (t)\rangle_{\rm ac}=0$,  and a second moment that is always larger than that in equilibrium
\begin{equation}\label{eq:2M}
    \langle x^2\vert \Lambda (t)\rangle_{\rm ac}= \frac{k_B T}{\kappa(t)} + \dfrac{\omega_a \gamma}{2 \kappa(t)}\sigma^2_a(t).
\end{equation}
Yet, the kick amplitude distribution being Gaussian, the accompanying distribution cannot be merely described by a Gaussian at a higher (effective) temperature. This can be demonstrated by the fact that, although the skewness of $P_{\rm ac}$ vanishes, its excess kurtosis
\begin{equation}
    \label{eq:Kex}
    \displaystyle\mathcal{K}_{\rm ex}(t)=\frac{3}{4}\omega_a\frac{\gamma}{\kappa(t)}\left[\frac{\sigma_a^2(t)}{\langle x^2\vert \Lambda (t)\rangle_{\rm ac}}\right]^2
\end{equation}
is nonzero whenever $\sigma_a^2(t)>0$ 
indicating in general the presence of fat tails. 
We recall readers that the equilibrium distribution, retrieved by taking the limit $\omega_a\to 0$ in Eq.~\eqref{eqn:ps_noneqsteady}, yielding a Gaussian distribution  given by
\begin{equation}
\label{eqn:P_eq}
P^{\rm eq}(x\vert \Lambda(t))=  \sqrt{\dfrac{\kappa(t)}{2\pi k_BT(t)}} \exp \left(-\frac{\kappa(t) x^2}{2k_{B}T(t)}\right).
\end{equation}
In the subsequent sections, having at hand the analytical expression of the instantaneous  $P$ 
(Eq.~\eqref{eqn:FokkerPlanck_solution}), accompanying $P^{\rm ac}$ (Eq.~\eqref{eqn:ps_noneqsteady}) and equilibrium $P^{\rm eq}$ (Eq.~\eqref{eqn:P_eq}) distributions, we will study in  detail the energetics and efficiency of  heat engines in the presence of PSN.

\section{Energetics and efficiency of active heat engines}
\label{stocengergy}
In exploring active systems within the realm of stochastic thermodynamics, we employ the definitions of work and heat along a single trajectory to study the stochastic energetics associated with the Langevin equations, following Sekimoto's formalism~\cite{Sekimoto98}. 
The stochastic  heat $q$ absorbed by the engine and the stochastic  work $w$  extracted from the engine, both associated with a specific stochastic trajectory read respectively
\begin{eqnarray}
 q&=&\int_0^\tau \frac{\partial U(x(t),t)}{\partial x}\circ \dot{x}(t)dt, \\
   w&=&-\int_0^\tau \frac{\partial U(x(t),t)}{\partial t}dt,
\label{eq:active_heat}
\end{eqnarray} 
where $\circ$ denotes the Stratonovich product.
Note that $q-w=U(x(\tau),\tau)-U(x(0),0)$ thus retrieving the first law of thermodynamics at the level of single trajectories.  Throughout the paper, we adopt the convention according to which positive (negative) heat signifies heat absorbed (dissipated) by the system from (to) the environment, and positive (negative) work  indicates work extracted (exerted) from (to) the system. 
The averages of heat and work accumulated
over one driving cycle of duration $\tau$ are defined as $Q=\langle q\rangle$ and
$W=\langle w\rangle$, where $\langle\cdot\rangle$ denotes an ensemble average over many realizations of the cyclic protocol driving the engine.
A customary way to characterize thermodynamic properties of stochastic heat engines is to focus on thermodynamic quantities averaged over many realizations of the engine's cycle, namely the average    heat absorbed from the hot bath $Q_h$, the average heat absorbed from the cold bath $Q_c$ and the average work extracted over all the cycle $W$. The `traditional' definition for  efficiency is given by
\begin{equation}
\eta = \frac{W}{Q_h}.
\label{eq:pseudo_efficiency}
\end{equation}
For engines put in contact with two  heat baths at equilibrium with temperatures $T_c<T_h$, the traditional definition of efficiency, Eq.~\eqref{eq:pseudo_efficiency} is such that, by virtue of the first law  applied to cycles $Q_h + Q_c - W =0$ and the second law of thermodynamics  $-\beta _h Q_h-\beta_c Q_c\geq 0$, where $ \beta_h = 1 / k_B T_h$, $\beta_c = 1 / k_B T_c$,  one retrieves the Carnot bound
\begin{equation}
    \eta\leq \eta_C=1-\frac{\beta_h}{\beta_c}.\label{eq:20}
\end{equation}
For heat engines embedded in active (or more generally, nonequilibrium) reservoirs, it is mandatory to include entropy flow contributions  $\Delta S_{\rm act}$ between the system and  the active (nonequilibrium) bath/s when formulating the second law of thermodynamics. For the simple scenario in which a heat engine is in contact with both equilibrium and active baths which do not interact between each other, one can decouple the environmental entropy flows in the total entropy production $\Delta S_{\rm tot}$ as 
\begin{equation}\label{eq:2LC}
    \Delta S_{\rm tot} = \Delta S_{\rm act}+Q_h/T_h+Q_c/T_c\geq 0.
\end{equation} Thus, combining the second law $\Delta S_{\rm tot}\geq 0 $ (Eq.~\eqref{eq:2LC}) together with  the first law  for cycles $W-Q_h - Q_c=0$, does not ensure that the traditional definition of efficiency $\eta$ given by Eq.~\eqref{eq:pseudo_efficiency} is bounded by the Carnot value.

An alternative framework for active heat engines was proposed in by Datta et al. in Ref.~\cite{Barato} by treating the source/s of activity as external, non-conservative forces that make the statistics of the engine to be genuinely out of equilibrium. In other words, it is not possible to describe the statistics of the engine using equilibrium (Boltzmann) distribution even when driven quasistatically. For such non-equilibrium conditions Datta et al. in Ref.~\cite{Barato}  focuses on the average of the so-called excess entropy production \cite{Onno, Hatano, Esposito2010}, denoted by $ \Delta S_{\rm ex}= \Delta S_{\rm tot}- \Delta S_{\rm hk}$, which is the average entropy produced atop the `housekeeping' $\Delta S_{\rm hk}\geq 0$ value that is the one produced in the quasistatic limit ($\Delta S_{\rm hk}= 0$ when only conservative forces are exerted on the system). The excess entropy obeys a second-law-like inequality for generic nonequilibrium processes, which, when applied to heat engines reads
\begin{eqnarray}
  \Delta S_{\rm ex}/k_{B}
= \mathcal{I}+(\beta_c-\beta_h)Q_h-\beta_c W \geq 0  ,\label{eq:datta}
\end{eqnarray}
where the quasistatic divergence $\mathcal{I}$ is an information-theoretic contribution that measures the `distance' between the statistics of the engine in the quasistatic limit to the equilibrium (Boltzmann) distribution. Notably, unlike $\Delta S_{\rm act}$, $\mathcal{I}$ is a purely statistical quantity that may be estimated without the need of knowing the underlying active mechanisms generating entropy flows between the system and the active baths; it quantifies the time-averaged distance between the accompanying distribution associated to the engine to the equilibrium Boltzmann distribution of its passive counterpart
\begin{equation}
\label{eq:I_inf}
\mathcal{I}=\int_{0}^{\tau} dt \int_{-\infty}^{\infty} dx\, P(x,t)\frac{d}{dt}\left[ \log{\frac{P^{\rm eq}(x\vert \Lambda(t))}{P^{\rm ac}(x\vert \Lambda(t))}}\right] .
\end{equation}
Here, $\tau$ is the  cycle time and $P(x,t)$, $P^{\rm ac}(x|\Lambda(t))$ and $P^{\rm eq}(x|\Lambda(t))$ are the  instantaneous   (Eq.~\eqref{eqn:FokkerPlanck_solution}), accompanying  (Eq.~\eqref{eqn:ps_noneqsteady}) and equilibrium  (Eq.~\eqref{eqn:P_eq}) distributions discussed in Sec.~\ref{sec:Quantifying_eficiency_in_active_engine}. Put simply, $\mathcal{I}$ quantifies the time-averaged divergence between the  engine's statistics in the quasistatic limit with respect to a the continually changing equilibrium distribution. This is why we call $\mathcal{I}$  {\em  quasistatic divergence} throughout the paper. Here and further we will use $\log$ for natural logarithm.

From the second law for the excess entropy production, Eq.~\eqref{eq:datta}, one can introduce a definition of  (`active') efficiency suitable for stochastic heat engines (where $W>0$ and $Q_h>0$) in contact with non-equilibrium reservoirs. In particular, when \(\mathcal{I}+\beta_c\eta_C Q_h \geq 0\), Eq.~\eqref{eq:datta} implies, after some algebra,
\begin{equation}
\eta_a\equiv \frac{W}{Q_h + \left(\displaystyle\frac{\mathcal{I}}{\beta_c - \beta_h}\right)} \leq \eta_C.
\label{eta_I}
\end{equation}
We emphasize that Eq.~\eqref{eta_I} is not specific to the PSN
model considered in this work; it follows from the 
second law for the excess entropy production [Eq.~(\ref{eq:datta})] which holds for Markovian systems under the action of  generic external (non-conservative) forces. The specific details of 
active dynamics impacts  the model-specific shape of the accompanying distribution
$P^{\rm ac}$, and consequently ${\mathcal{I}}$. As discussed in Ref.~\cite{Barato}, a positive $\mathcal{I}>0$ permits that an active heat engine's traditional efficiency $\eta=W/Q_h$ exceeds the Carnot limit thus violating the inequality~\eqref{eq:20} while  Eq.~\eqref{eta_I} remains valid. In the passive limit, \(P^{\rm ac}(x\vert \Lambda(t))=P^{\rm eq}(x\vert \Lambda(t))\), which yields  \(\mathcal{I}=0\), and the inequality~\eqref{eta_I}  reduces to Carnot bound~\eqref{eq:20}. 

Note that the fact that $\mathcal{I}$ is {\em not} a Kullback-Leibler divergence implies that it can take both positive and negative values. Thus, one may find parameter values for which a system behaves like an active heat engine (with $W>0$ and $Q_h>0$) yet with negative $\mathcal{I}\leq 0$ as long as it is not so negative as to render the denominator of Eq.~\eqref{eta_I} negative.
On the other hand, when \(\mathcal{I}+\beta_c\eta_C Q_h \leq 0\) the device operates as a refrigerator or heat pump according to the signs of $Q_c$ and $W$.
In the refrigerator regime \cite{Baratoref26},
\begin{equation}
Q_c>0,\qquad Q_h<0,\qquad W<0,
\end{equation}
i.e., heat is extracted from the cold bath and work is supplied to the system.
Substituting $Q_h=W-Q_c$ gives
\begin{equation}
\mathrm{COP_a}\equiv \frac{Q_c}{(-W)-\mathcal{I}/\beta_h}\le \mathrm{COP}_C,
\label{eq:COP_active}
\end{equation}
provided the denominator is positive, $(-W)-\mathcal{I}/\beta_h>0$, with 
\begin{equation}
\mathrm{COP}_C \equiv \frac{\beta_h}{\beta_c-\beta_h}.
\label{eq:COP_Carnot}
\end{equation}
 the standard (Carnot) coefficient of performance \cite{Rana2016}. 
 
In this paper, we study an active (PSN) heat engine model with parameter values ensuring  work extraction  $W>0$ and heat absorption from the hot thermal bath $Q_h>0$ while satisfying  \(\mathcal{I}+\beta_c\eta_C Q_h \geq 0\). The simplicity of the PSN approach enables us to explore both positive and negative values of $\mathcal{I}$, with the latter not being reported yet in the literature~\cite{Barato}.
To characterize the engine performance, we consider in the next section active heat engines  put in contact with equilibrium (thermal) and nonequilibrium (PSN noise) baths at different temperatures. For simplicity, in Sec.~\ref{sec:Two_stroke_engine} we begin with the study of engines run by  two-stroke cyclic protocols for the stiffness and for the temperature of the (equilibrium and nonequilibrium) baths, and relegate for the Appendix~\ref{app:stirling_protocol} a follow-up analysis for cyclic protocol analogous to that of a Stirling engine.
\begin{figure}[h]
\includegraphics[width=0.35\textwidth]{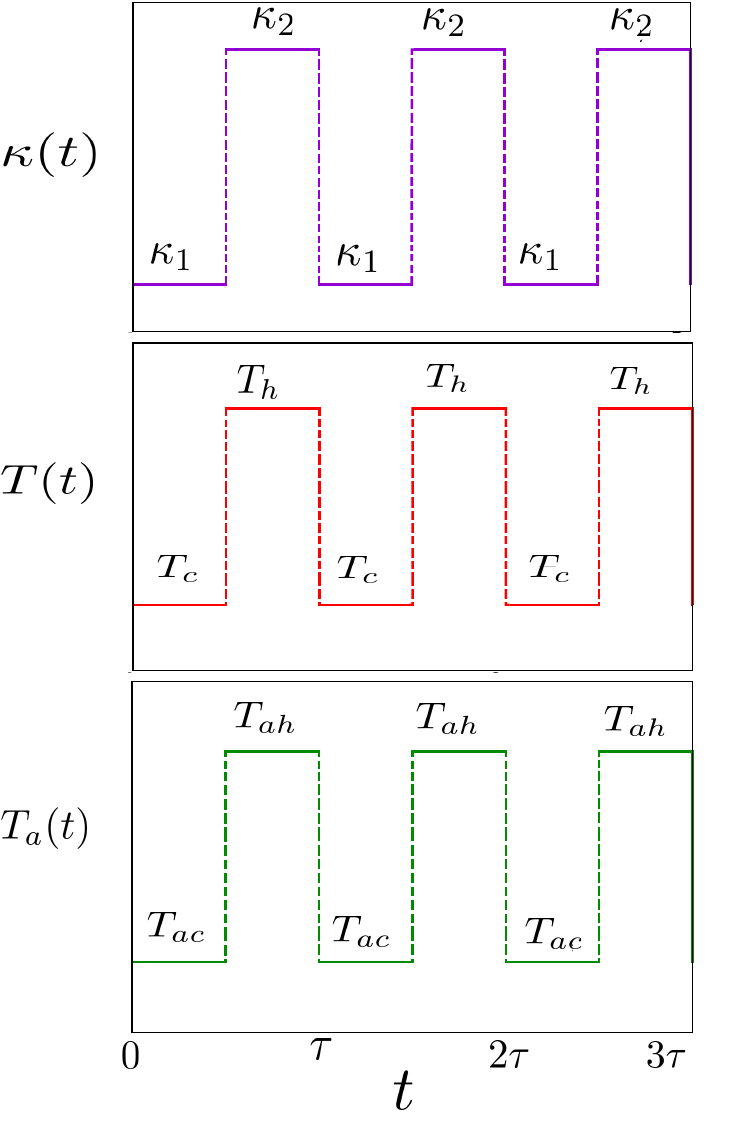}\hfill
\caption{Sketch of the protocol associated with the two-stroke active heat engine with Poisson shot noise (PSN): stiffness $\kappa(t)$ (top panel), temperature of the thermal bath $T(t)$ (middle panel),  and  effective temperature of the PSN  $T_{a}(t)$ (bottom panel) as a function of time $t$ over three  cycle periods, see Eq.~(\ref{maximum_efficiency_protocol}). Here, $\kappa_1$ and $\kappa_2$ are the minimum and maximum values of the trap stiffness. Similarly, $T_c$ and $T_h>T_c$ denote the temperature of the cold and hot thermal baths respectively, while  $T_{ac}$ and $T_{ah}>T_{ac}$  the (effective) temperature of the PSN active baths, see Eq.~\eqref{eq:Tadef}. In our simulations, we repeated the protocol cyclically over periods of duration $\tau$.}
\label{fig:protocol_psn}   
\end{figure}  
\section{Two-stroke PSN heat engine: protocol and analytical results}
\label{sec:Two_stroke_engine}{In the numerical analysis that follows, we take piconewtons ($\mathrm{pN}$), nanometers ($\mathrm{nm}$), and seconds ($\mathrm{s}$), as the units of force, length, and time, respectively. Thus, position is expressed in $\mathrm{nm}$, trap stiffness in $\mathrm{pN/nm}$, friction coefficient in $\mathrm{pN\,s/nm}$, time quantities in $\mathrm{s}$, and rates in $\mathrm{s^{-1}}$. Thermal and active temperatures are measured in Kelvin. In this Section, we consider a two-stroke engine protocol and apply the theory developed above to study this specific example. Rather than focusing on a specific experimental system, all our simulation results reported in our work are intended as a proof of principle for generic active systems.
\subsection{Engine protocol}
\label{sec:Solving the Langevin equation} Following~\cite{Edgar24}, we reparameterize the PSN variance as
\begin{equation}
    \sigma^2_a(t) = \tau_a\frac{k_{\rm B}T_a(t)}{\gamma},\label{eq:6}
\end{equation}
with $\tau_a$ being a parameter with dimension of time, and $T_a$ an effective (`active') temperature that is defined, from Eq.~\eqref{eq:6} as,
\begin{equation}\label{eq:Tadef}
    T_a(t)= \frac{\gamma }
    {\tau_a k_{\rm B}}\sigma^2_a(t).
\end{equation}
For a better clarity, let us rewrite the Langevin equation~\eqref{eqn:Langevin_2} in terms of these new parameters
\begin{equation}\label{eq:LEPSN}
   \gamma \dot{x}(t)=-\kappa(t)x(t)+\sqrt{2k_BT(t)\gamma}\xi(t)+\sqrt{k_{\rm B}  T_a(t)\gamma\omega_a\tau_a}\xi_a(t),
\end{equation}
with $\xi_a(t)$ a zero-mean $\langle\xi_a(t)\rangle=0$ unit variance $\langle\xi_a(t)\xi_a(t')\rangle=\delta(t-t')$ PSN with shot rate $\omega_a$. We consider  $\kappa(t)$,  $T(t)$ and $T_a(t)$ as step-like time-dependent (see Fig. \ref{fig:protocol_psn}.) that take on two different values:
\begin{eqnarray}
    \kappa(t) = 
\begin{cases}
\kappa_{1} & 0 \leq t \leq \tau/2\\
\kappa_{2} & \tau/2 < t \leq \tau
\end{cases} 
\nonumber
,\\
T(t) = 
\begin{cases}
T_{c} & 0 \leq t \leq \tau/2\\
T_{h} & \tau/2 < t \leq \tau
\end{cases}
,\label{maximum_efficiency_protocol}\\
T_{a}(t) = 
\begin{cases}
T_{ac} & 0 \leq t \leq \tau/2\\
T_{ah} & \tau/2 < t \leq \tau.
\end{cases}\nonumber
\end{eqnarray}  
This way, the engine's protocol consists of two fast strokes interspersed by relaxations. The stroke $\kappa_1 \to \kappa_2>\kappa_1$ is analogous to a  compression accompanied by a heating of the baths ($T_c\to T_h$, and $T_{ac}\to T_{ah}$). On the other hand, the stroke $\kappa_2 \to \kappa_1$ is analogous to an expansion accompanied by a cooling of the baths ($T_h\to T_c$, and $T_{ah}\to T_{ac}$).  As we show later, this specific form of the protocols ensure that most of the thermodynamic quantities can be calculated  analytically, at least in the large cycle time $\tau$ limit.
\begin{figure*}[!t]
\centering
\includegraphics[width=0.95\textwidth,
trim=10 10 0 0,
clip]{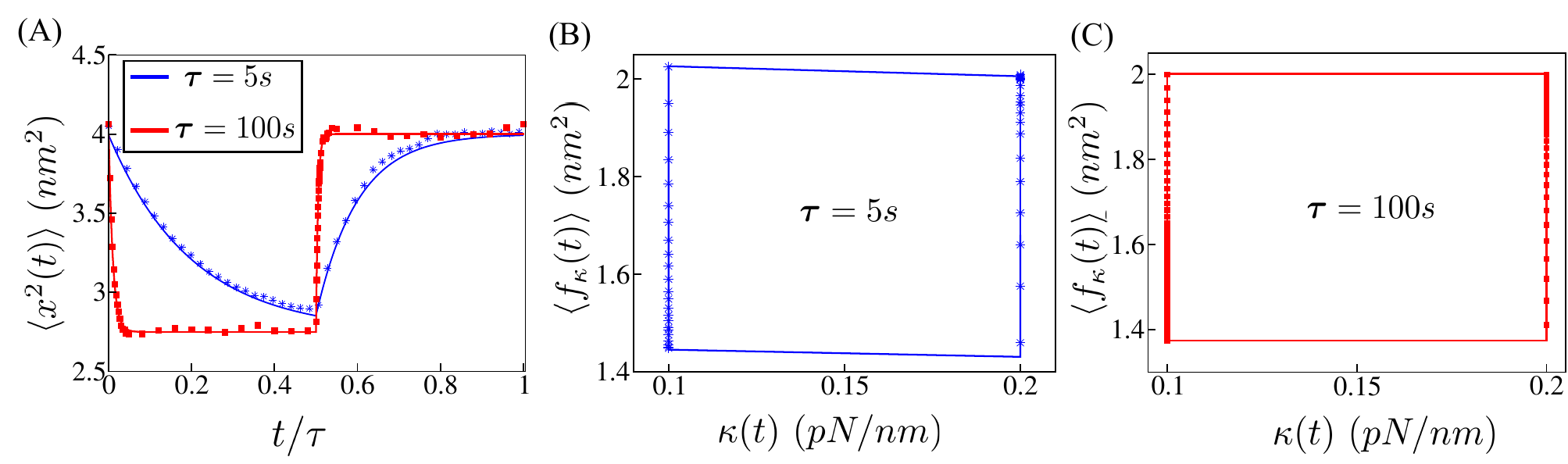}\hfill
\caption{Second moment of the particle position and Clapeyron diagram for the two-stroke active heat engine with Poisson shot noise (PSN) near and far from the quasistatic limit. (A) Variance $\langle x^2(t)\rangle$ of the particle position  as a function of time  (rescaled by the cycle time) $t/\tau$, for cycle times $\tau=5\,\mathrm{s}$ and $\tau=100\,\mathrm{s}$. (B,C) Clapeyron diagrams in the $\langle f_\kappa(t)\rangle$-$\kappa(t)$ plane, with $f_k(t)= x^2/2$ the generalized force conjugate to $\kappa$; the enclosed area gives the average work per cycle.
In all panels, results from numerical simulations are displayed with symbols, while analytical results (given  by Eqs.~\eqref{eq:xsqr1} and~\eqref{eq:xsqr2}) in solid lines. Parameters are $\gamma=0.2\,\mathrm{pN\,s/nm}$, $T_c=14.5\,\mathrm{K}$, $T_h=43.5\,\mathrm{K}$, $\kappa_1=0.1\,\mathrm{pN/nm}$, $\kappa_2=0.2\,\mathrm{pN/nm}$,  $\tau_a=1.0\,\mathrm{s}$, $T_{ac}=21.7\,\mathrm{K}$, $T_{ah}=58.0\,\mathrm{K}$, $\omega_a=0.5\,\mathrm{s}^{-1}$, number of simulations  $10^5$ and simulation time step $dt=0.1~\mathrm{ms}$.}
\label{Fig: Cycle_PC}
\end{figure*}

We first analyze the engine dynamics over a full cycle in both limits of short and long cycle times $\tau$ as compared with the relaxation times $\gamma/\kappa_1$ and $\gamma/\kappa_2$ associated with the two values of the stiffness. To this aim, we focus on the second (i.e. the first non-zero)  moment $\langle x^2(t)\rangle$ vs time $t$ during a cycle. Fig.~\ref{Fig: Cycle_PC}.~(A) shows an excellent agreement between results  from numerical simulations and analytical calculations, revealing that for $\tau$ large (around $100$ times the relaxation times) the second moment exhibits a step-like behavior, indicating that the particle has sufficient time to reach the two values of the accompanying density during the protocol. On the other hand, when $\tau$ is of the same order of magnitude as the relaxation times, the second moment exhibits a slow relaxation behaviour that is not able to reach the accompanying distribution at any time of the cycle. A further characterization of these relaxation properties is through the Clapeyron diagram associated with this protocol. In particular, the  work exerted in $[t,t+dt]$ averaged over many realization with this protocol takes the form
\begin{equation}
    dW(t)= -\frac{ x^2(t) }{2} d\kappa (t).
    \label{eqn:work_general}
\end{equation}
Here, we identify $f_\kappa (t)=\partial U/\partial\kappa  =   x^2(t)/2$ as the 
generalized force, and $d\kappa (t)$ its conjugate generalized displacement. This motivates   our choice to build the $f_\kappa$ vs $\kappa$ Clapeyron-like diagram, shown in Fig.~\ref{Fig: Cycle_PC}.~(B) and Fig.~\ref{Fig: Cycle_PC}.~(C)  for $\tau$ small and $\tau$ large as compared with the relaxation times. The deviation from the quasistatic limit at small $\tau$ is characterized by a time-inhomogeneous speed at which the Clapeyron diagram is traveled.  
We  focus below and in the rest of the paper on the quasistatic, large $\tau$  limit.

In the quasistatic limit, the particle  relaxes to the stationary state at all times within  each step of the piecewise-constant driving protocol. Therefore, one can assume in good (quasistatic) approximation that
 \begin{equation}
     P(x,t)\simeq P^{\rm ac}(x\vert \Lambda(t)),
 \end{equation}  such that analytical calculations can be tackled using the stationary distributions associated with the instantaneous values of the control parameters.  Fig.~\ref{Fig: distribution}. shows, the probability density $P^{\rm ac}(x|\Lambda)$ during the compression and expansion branches for a cycle time much larger than the relaxation time. 
 Although the kick amplitude distribution $\rho_a(y)$ is Gaussian, the  accompanying density inherits  (non-Boltzmann) features as a result of the interaction with the active PSN~\cite{Edgar24}.   The resulting accompanying density (given by Eq.~\eqref{eqn:ps_noneqsteady}) exhibits pronounced non-Gaussian tails that are absent in the equilibrium reference distribution $P^{\rm eq}(x|\Lambda)$~Eq.~\eqref{eqn:P_eq}, reflecting the nonequilibrium nature of the PSN fluctuations. These non-Gaussian accompanying distributions provide the basis for the analytical results discussed below in the quasistatic limit.

\begin{figure*}[!htbp]
\includegraphics[width=0.8\textwidth, trim=1 10 0 0, clip]{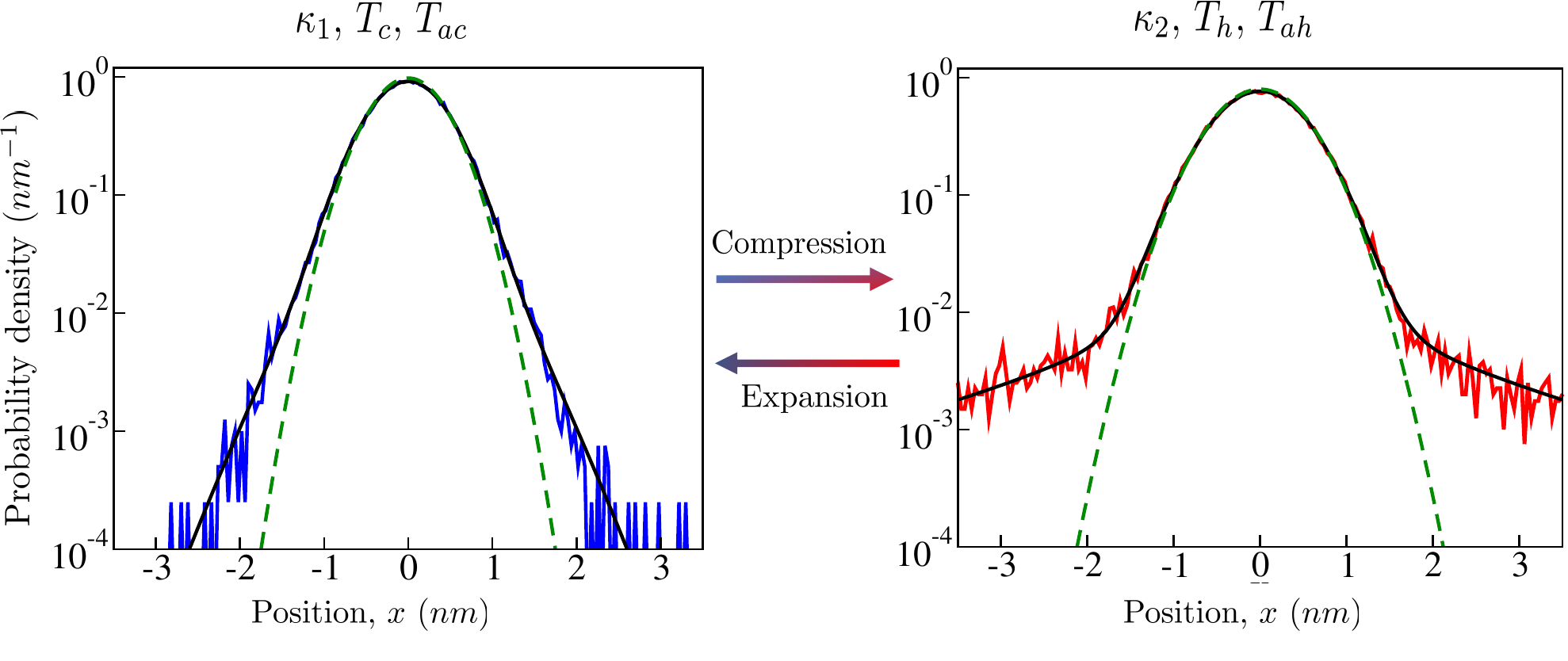}
\caption{Probability density of particle position in the two-stroke heat engine with Poisson shot noise~(PSN) obtained from numerical simulations and analytical calculations:
accompanying distribution $P^{\rm ac}(x\vert\Lambda)$ associated with $\Lambda_1=(\kappa_1,T_c,T_{ac})$ (blue line) and with $\Lambda_2=(\kappa_2,T_h,T_{ah})$ (red line), and equilibrium distribution obtained for $\omega_a=0$ $P^{\rm eq}(x\vert\Lambda)$ (Eq.~\eqref{eqn:P_eq}, green dashed line for $\Lambda_1$ and $\Lambda_2$).
Parameter values are $\kappa_1=0.6\,\mathrm{pN/nm}$, $T_c=7.25\,\mathrm{K}$, $T_{ac}=7.25\,\mathrm{K}$ in the cold isothermal, and 
$\kappa_2=4.0\,\mathrm{pN/nm}$, $T_h=72.5\,\mathrm{K}$,  $T_{ah}=362.3\,\mathrm{K}$, in the hot isothermal.
The roughed blue and red lines are normalized histograms obtained from numerical simulations, and the black lines in both panels are obtained using Eq.~\eqref{eqn:ps_noneqsteady}. 
Other simulation parameters  $\gamma=0.2\,\mathrm{pN\,s/nm}$, $\tau_a=1.0\,\mathrm{s}$, $\omega_a=0.5\,\mathrm{s}^{-1}$, cycle time $\tau=100\,\mathrm{s}$, number of simulations  $10^5$ and simulation time step $dt=0.1\,\mathrm{ms}$.
\label{Fig: distribution}}
\end{figure*}

\subsection{Performance in the quasistatic limit: analytical expressions}
\label{subsec:B}
In this subsection, we evaluate key thermodynamic quantities characterizing the two-stroke engine's performance in  the quasistatic limit using the concepts developed in Sec. \ref{sec:Quantifying_eficiency_in_active_engine}. The two-stroke protocol has the main advantage of its simplicity regarding thermodynamic analyses, as all the heat and work exchanges may be expressed directly as internal energy changes. The two strokes $\Lambda_1\to\Lambda_2$  and $\Lambda_2\to\Lambda_1$  are instantaneous: there is no heat exchange with the baths and the work equals to the internal energy change. On the other hand, within the relaxations at $\Lambda_1$ and $\Lambda_2$, the stiffness is kept constant which implies zero work, that is the heat dissipated equals to the internal energy change. 
Since each step of the two-stroke engine  are relaxation processes with zero work cost $W_c=0$ and $W_h=0$, the average heat dissipated in the cold isothermal equals to the average change in internal energy
\begin{eqnarray}\label{eq:31}
Q_c=\frac{\kappa_1}{2}\left[\langle x^2(\tau/2^-)\rangle-\langle x^2(0)\rangle\right],
\end{eqnarray}
and similarly for the average heat dissipated in the hot isothermal
\begin{equation}
\label{eq:heat_active_passivea}
Q_{h} = \frac{\kappa_2}{2}\left[
\langle x^2(\tau^-)\rangle
-\langle x^2(\tau/2^+)\rangle
\right].
\end{equation}
Similarly, the average work per cycle may be found by summing the internal energy change  in the two instantaneous switches $\Lambda_1\to\Lambda_2$  and $\Lambda_2\to\Lambda_1$: 
\begin{equation}\label{eq:33}
W = \frac{(\kappa_2-\kappa_1)}{2}
\left[\langle x^2(\tau^+)\rangle-\langle x^{2}(\tau/2^-)\rangle\right].
\end{equation}
In the above equations, the superscripts ``$-$'' and ``$+$'' denote the values immediately before and immediately after an instantaneous switch of the protocol parameters, respectively. Eqs.~(\ref{eq:31}-\ref{eq:33}) reveal that the key thermodynamic quantities depend solely on the second moment of the particle's position. Analytical expressions for average heat and work flows, as well as the engine's efficiency are readily available by employing the analytical formulae for the second moment at all times that we derive in Appendix~\ref{app:two_step_protocol}.  
More precisely, we may use the statistics of the accompanying distributions to get, for example,
\begin{equation}
\label{eq:heat_active_passivea2}
Q_{h} = \frac{\kappa_2}{2}\left[
\langle x^2\vert \Lambda_2\rangle_{\rm ac}
-\langle x^2\vert \Lambda_1\rangle_{\rm ac}
\right],
\end{equation}
and
\begin{equation}\label{eq:332}
W = \frac{(\kappa_2-\kappa_1)}{2}
\left[
\langle x^2\vert \Lambda_2\rangle_{\rm ac}
-\langle x^2\vert \Lambda_1\rangle_{\rm ac}
\right].
\end{equation}
Using Eq.~\eqref{eq:2M}, we evaluate the averages $\langle x^2\vert \Lambda_1\rangle_{\rm ac}$ and $\langle x^2\vert \Lambda_2\rangle_{\rm ac}$ yielding to analytical formulae for $Q_h$ and $W$ valid for all parameter choices of $\Lambda_1$  and $\Lambda_2$. In particular,  
the heat $Q_h$ may be decomposed in the sum of two independent (thermal $Q_{\rm th}$, and active $Q_{\rm a}$) contributions 
\begin{align}
\label{eq:heat_active_passive}
Q_{h} &= Q_{\rm th}+Q_{\rm a} ,
\end{align}
where
\begin{align}
Q_{\rm th} &=
\frac{k_B T_h}{2}
\left(1-\frac{\kappa_2}{\kappa_1}\frac{T_c}{T_h}\right),
\\
Q_{\rm a} &=\omega_a\tau_a\dfrac{k_B T_{ah}}{4}\left(1-\dfrac{\kappa_2}{\kappa_1}{\dfrac{T_{ac}}{T_{ah}}}\right).\label{eq:Qaa}
\end{align}
Analogously, the average work per cycle may be decomposed as the sum of two independent (thermal $W_{\rm th}$, and active $W_{\rm a}$) contributions 
\begin{align}
\label{eq:work_thermal_act}
W &= W_{\rm th}+W_{\rm a}, 
\end{align}
where
\begin{align}
\label{eq:work_thermal_act2}
W_{\rm th} &=
\frac{k_{B}T_h}{2}\left(1-\frac{\kappa_1}{\kappa_2}\right)
\left(1-\frac{\kappa_2}{\kappa_1}\frac{T_{c}}{T_h}\right),
\\
W_{\rm a} &=
\omega_a\tau_a\dfrac{k_{B}T_{ah} }{4}\left(1-\frac{\kappa_1}{\kappa_2}\right)
\left(1-\frac{\kappa_2}{\kappa_1}\frac{T_{ac}}{T_{ah}}\right).\label{eq:wa}
\end{align}
Eqs.~(\ref{eq:heat_active_passive}-\ref{eq:wa}) reveal that, in the absence of (active) PSN, i.e. for $\omega_a\tau_a=0$, the requirement for both $W=W_{\rm th}$ and $Q_h=Q_{\rm th}$ to be positive (engine behaviour), implies that $\kappa_2/\kappa_1\leq T_h/T_c$. Thus the 'traditional' efficiency in that case is bounded by the Carnot limit $\eta=W_{\rm th}/Q_{\rm th}=1-(\kappa_1/\kappa_2)\leq \eta_C$. 

\subsection{Quasistatic divergence and active efficiency}
Less intuitive yet very insightful is the fact that, in the presence of PSN  with generic parameter values, the `traditional' efficiency in the quasistatic limit reads 
\begin{equation}
    \eta=\frac{W}{Q_h}=1-\frac{\kappa_1}{\kappa_2}.
    \label{eq:pseudoeffi}
\end{equation}
The fact that the `traditional'  efficiency takes the same value with or without PSN challenges its validity to properly characterize the engine's performance. In particular, as we will show later, one may  find parameter values for which  $\eta$ in Eq.~\eqref{eq:pseudoeffi} violates the Carnot bound yet the machine functioning as an engine $W>0$ and $Q_h>0$---a result unattainable in the absence of active noise. 
To resolve this conundrum,  we focus on the active efficiency defined in Eq.~\eqref{eta_I},  and rewritten  here for convenience as
\begin{equation}
\eta_a =
\frac{\eta}{
1+\dfrac{\mathcal{I}}{Q_h(\beta_c-\beta_h)}
},
\label{eta_I2}
\end{equation}
with $\eta$ given by Eq.~\eqref{eq:pseudoeffi}, and where we recall $\beta_c$ and $\beta_h$ are defined in terms of the temperatures of the thermal baths. As we will show later, $\eta_a$ defined by Eq.~\eqref{eta_I2} 
 is compliant with the Carnot limit $\eta_a\leq \eta_C$ for any parameter choices where the machine works as an engine.  The key advantage of using the two-stroke protocol is that in the quasistatic limit, the quasistatic divergence $\mathcal{I}$ may be evaluated semi-analytically as a combination of six Kullback-Leibler divergences
\begin{widetext}
\begin{eqnarray}
\mathcal{I} &=& \int_{-\infty}^{\infty}dx [P^{\rm ac}(x\vert \Lambda_2)-P^{\rm ac}(x\vert \Lambda_1)]\left(\log{\dfrac{P^{\rm ac}(x\vert \Lambda_2)}{P^{\rm eq}(x\vert \Lambda_2)}}-\log{\dfrac{P^{\rm ac}(x\vert \Lambda_1)}{P^{\rm eq}(x\vert \Lambda1)}}\right),\nonumber\\
\nonumber\\
&=& D[P^{\rm ac}(x\vert \Lambda_2)||P^{\rm eq}(x\vert \Lambda_2)]+D[P^{\rm ac}(x\vert \Lambda_1)||P^{\rm eq}(x\vert \Lambda_1)]-D[P^{\rm ac}(x\vert \Lambda_2)||P^{\rm eq}(x\vert \Lambda_1)]-D[P^{\rm ac}(x\vert \Lambda_1)||P^{\rm eq}(x\vert \Lambda_2)]
\nonumber \\ 
\nonumber\\
&+&D[P^{\rm ac}(x\vert \Lambda_2)||P^{\rm ac}(x\vert \Lambda_1)]+D[P^{\rm ac}(x\vert \Lambda_1)||P^{\rm ac}(x\vert \Lambda_2)].
\label{I_general_psn}
\end{eqnarray} 
\end{widetext}
Here, \( P^{\rm ac}(x\vert \Lambda_1) \) and \( P^{\rm eq}(x\vert \Lambda_1) \) denote respectively the accompanying and equilibrium densities associated with the position $x$ during \( 0 < t \leq \tau/2 \), while \( P^{\rm ac}(x\vert \Lambda_2) \) and \( P^{\rm eq}(x\vert \Lambda_2) \)  the accompanying and equilibrium densities during \( \tau/2 < t \leq \tau \). Note that in the zero PSN limit where $P^{\rm ac}=P^{\rm eq}$, one gets $\mathcal{I}=0$ as expected.

To understand how the quasistatic divergence accumulates during a cycle, we introduce the cumulative quantity in the interval $[0,t]$ defined as,
\begin{equation}
 I(t)=
\int_0^t ds
\left\langle
\frac{d}{ds}
\log\frac{P^{\rm eq}(x|\Lambda(s))}
{P^{\rm ac}(x|\Lambda(s))}
\right\rangle,
\label{eq:I_cumulative}
\end{equation}
where, the average is taken over the instantaneous distribution $P(x,s)$. By definition, $\mathcal I= I(\tau)$.
For the two-stroke protocol, the control parameters remain constant within each half-cycle and change only at the two switching events, $\Lambda_1\rightarrow\Lambda_2$ at $t=\tau/2$ and $\Lambda_2\rightarrow\Lambda_1$ at $t=\tau$. Therefore, $I(t)$ remains constant during the relaxations and changes only at the switching events.

The total quasistatic divergence accumulated over one cycle can thus be written as the sum of the contributions from the compression and expansion switches,
\begin{equation}
\mathcal I=I_c+I_e,
\label{eq:I_two}
\end{equation}
where,
\begin{equation}
    I_c = \left\langle \log \frac{P^{\rm eq}(x\vert\Lambda_2)}{P^{\rm eq}(x\vert\Lambda_1)}-\log \frac{P^{\rm ac}(x\vert\Lambda_2)}{P^{\rm ac}(x\vert\Lambda_1)}\right\rangle_{P(x\vert\Lambda_1)},\label{eq:50}
\end{equation}
and
\begin{equation}
    I_e = \left\langle \log \frac{P^{\rm eq}(x\vert\Lambda_1)}{P^{\rm eq}(x\vert\Lambda_ 2)}-\log \frac{P^{\rm ac}(x\vert\Lambda_1)}{P^{\rm ac}(x\vert\Lambda_2)}\right\rangle_{P(x\vert\Lambda_2)}.\label{eq:52}
\end{equation}
Here, we use the shorthand notation $P(x|\Lambda_1)=P(x,\tau/2^-)$ and $P(x|\Lambda_2)=P(x,\tau^-)$, corresponding to the distributions immediately before the two switching events. In the quasistatic limit, these distributions approach $P^{\rm ac}(x|\Lambda_1)$ and $P^{\rm ac}(x|\Lambda_2)$, respectively, recovering Eq.~(\ref{I_general_psn}).
Figure~\ref{fig:cumulative_I} illustrates, this behavior for two representative parameter sets corresponding to negative and positive values of \(\mathcal I\). In Fig.~\ref{fig:cumulative_I}~(A), the first switch contributes positively, while the second switch produces a larger negative contribution, resulting in \(\mathcal I<0\). Conversely, in Fig.~\ref{fig:cumulative_I}.~(B), the first switch gives a negative contribution, while the second produces a larger positive contribution, leading to \(\mathcal I>0\). Thus, the sign of the quasistatic divergence is determined by the balance between the contributions from the two switching events. These contributions depend on the specific shape of the accompanying  distributions and their corresponding equilibrium reference distributions. To further understand how these distributions determine the sign of $\mathcal{I}$, we examine the spatial contributions to the integral in Eq.~(\ref{I_general_psn}) for a representative parameter set in Appendix \ref{app:twostroke_add}.
\begin{figure}[htpb!]
\hspace{-0.5cm}
 \includegraphics[width=0.43\textwidth]{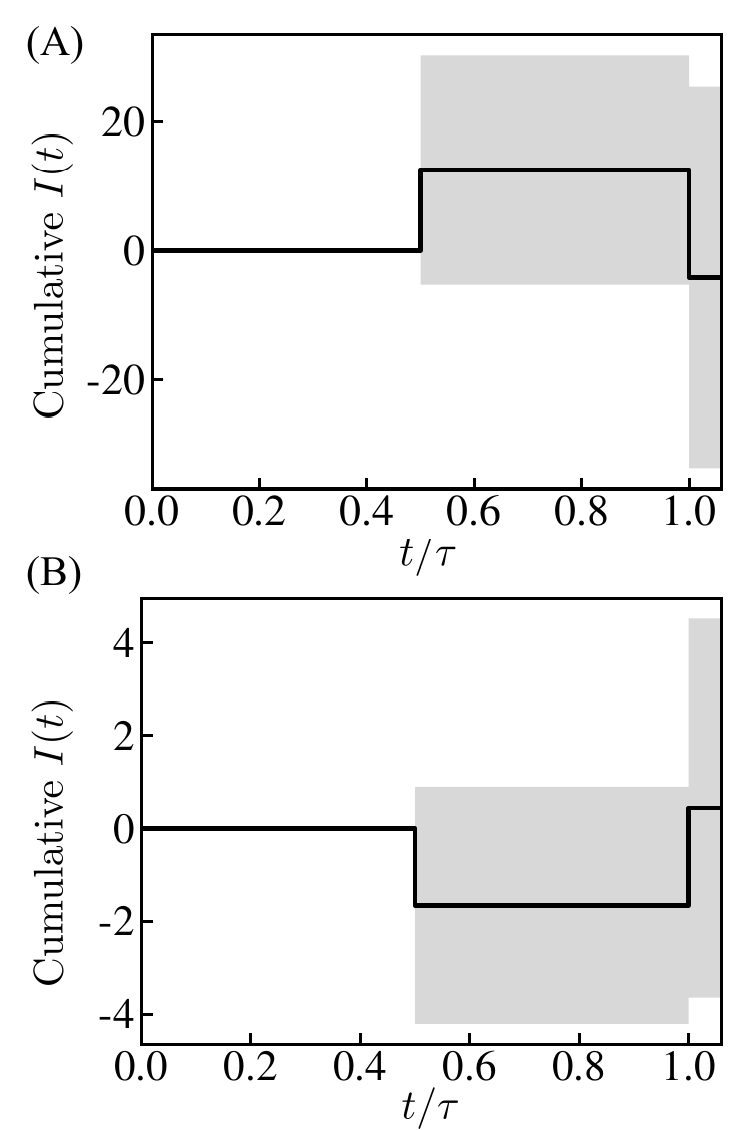}\hfill
\caption{Cumulated quasistatic divergence $I(t)$  up to time $t$ [see Eq.~\eqref{eq:I_cumulative}] during one cycle of the two-stroke active heat engine estimated from numerical simulations (black line). The grey shaded regions are standard deviations around the average value, illustrating the trajectory
fluctuations of $I(t)$.
 Because 
$\Lambda(t)$ is constant within each half cycle, $I(t)$ changes
only at the two switches at $t=\tau/2$ (compression) and at $t=\tau$ (expansion). The two panels are two representative example parameter choices that lead to different signs of $\mathcal{I}=I(\tau)$. (A)   Example case where $\mathcal{I}\simeq -4.21<0$ at the end of the cycle, where parameters are $\kappa_1=0.1~{\rm pN/nm},~\kappa_2=0.2~{\rm pN/nm},~T_c=14.5~{\rm K},~T_h=43.5~{\rm K},~T_{ac}=21.7~{\rm K}$, and $T_{ah}=58.0~{\rm K}$.~(B)  Example case where $\mathcal{I}=0.44>0$ at the end of the cycle with parameters are $\kappa_1=0.1~{\rm pN/nm},~ \kappa_2=0.32~{\rm pN/nm},~T_c=T_{ac}=14.5~{\rm K},~ T_h=43.5~{\rm K}$, and $T_{ah}=58.0~{\rm K}$. In both panels the other parameters are $\gamma=0.2\,\mathrm{pN\,s/nm}$, $\tau_a=1.0\,\mathrm{s}$ and
 $\omega_a=100\,\mathrm{s}^{-1}$.}
\label{fig:cumulative_I}
\end{figure}

\section{Two-stroke PSN heat engine: numerical simulations}
\label{sec:Two_stroke_engine2}
The results developed in the previous section for the quasistatic limit, suggests to introduce the rescaled temperature and stiffness values 
\begin{equation}\label{eq:47}
    \overline{\kappa}\equiv \frac{\kappa_2}{\kappa_1},\quad \overline{T}=\frac{T_h}{T_c},\quad \overline{T_a}=\frac{T_{ah}}{T_{ac}},
\end{equation}
which enable a drastic simplification of the study of the engine's behavior for different parameter values. In particular, we may discuss the sign of the average heat and work values in Eqs.~(\ref{eq:heat_active_passive}-\ref{eq:wa})  by noticing the proportionality relations
\begin{eqnarray}
  Q_{\rm th } &\propto &\left(\overline{T}-\overline{\kappa}
    \right),\;\; Q_a \propto \left(\overline{T_a}-\overline{\kappa}\right)
\label{eq:48}\\
    W_{\rm th } &\propto &\left(\overline{\kappa}-1
    \right)\left(\overline{T}-\overline{\kappa}
    \right),\;\;   W_{a } \propto \left(\overline{\kappa}-1
    \right)\left(\overline{T_a}-\overline{\kappa}
    \right).\label{eq:49}
\end{eqnarray}
Relations~(\ref{eq:48}-\ref{eq:49}) establish that, although $\bar{\kappa}$, $\overline{T}$ and $\overline{T_a}$ are all greater or equal than one, the sign of the work extracted and heat uptake per cycle depends directly on the sign of the differences $\left(\overline{T}-\overline{\kappa}
    \right)$ and $\left(\overline{T_a}-\overline{\kappa}
    \right)$, that is on how large are the ratios of (bath and effective) temperatures with respect to the ratio of stiffnesses. We will exploit this rationale in the following by  performing numerical simulations, focusing on representative parameter values. 
  \begin{table}
    \begin{center}
\begin{tabular}{|c|c|c|c|c|c| }
\hline & $\bar{\kappa}$& $\overline{T}$ & $\overline{T_a}$ & $Q_{\rm th},\;W_{\rm th}$& $Q_{\rm a},\;W_{\rm a}$\\
\hline
 Sec.~\ref{sec:Va} & 2 & 4&  1 & + & --\\ \hline
 Sec.~\ref{sec:Vb} & 2 & 3&  2.6 & + & + \\ \hline
Sec.~\ref{sec:Vc} & 3.2  & 3  & 4 &  -   & +   \\  \hline
\end{tabular}
\end{center}
\caption{Values of the key rescaled parameters [see Eq.~\eqref{eq:47}] for the two-stroke machine associated with the key case studies developed in different subsections (see first column), illustrating its hybrid  operational mode. The last two columns indicate the sign of the thermal (fourth column) and active (fifth column) contributions to the averages of the heat uptake from the hot bath and of the work extracted per cycle, cf. Eqs.~(\ref{eq:48}-\ref{eq:49}). In other words `$+$' stands for engine-like operation and `$-$' for refrigeration.  Note that in all parameter sets, we explored the regime where the total heat and work $Q_h = Q_{\rm th}+Q_{\rm a}>0$ and $W_{\rm th}+W_{\rm a}>0$ i.e. an overall engine operation. \label{TableRita}}
\end{table}
The three sets of parameters for the rescaled temperature and stiffness are such that they enable us to inspect a rich spectrum of thermodynamic regimes, all ensuring $W>0$ and $Q_h>0$ for the total work and heat uptakes (thermal+PSN), see Table~\ref{TableRita}. In particular, our minimal model may operate simultaneously as heat engine and (with respect to the thermal/PSN baths) and as refrigerator (with respect to the PSN/thermal baths) 
\begin{itemize}
    \item In  Sec.~\ref{sec:Va}, we  tackle the case in which the machine behaves like an engine with respect to the thermal bath ($Q_{\rm th}>0$ and $W_{\rm th}>0$), yet as a refrigerator with respect to the PSN bath ($Q_{\rm a}<0$ and $W_{\rm a}<0$).
     \item In  Sec.~\ref{sec:Vb}, we  tackle the case in which the machine behaves like an engine with respect to the thermal bath ($Q_{\rm th}>0$ and $W_{\rm th}>0$), and also with respect to the PSN bath ($Q_{\rm a}>0$ and $W_{\rm a}>0$).
      \item In  Sec.~\ref{sec:Vc}, we  tackle the case in which the machine behaves like a refrigerator with respect to the thermal bath ($Q_{\rm th}<0$ and $W_{\rm th}<0$), yet as an engine with respect to the PSN bath ($Q_{\rm a}>0$ and $W_{\rm a}>0$).
\end{itemize}

\subsection{\texorpdfstring{Homogeneous activity ($\bar{\kappa}>1$, $\overline{T}>1$, $\overline{T_a}=1$): thermal  engine, active refrigerator}{text}}

\label{sec:Va}

As a first case study, we focus  in the simple scenario in which the two (active) PSN baths are identical, i.e. $\overline{T_a}=1$  while $\overline{\kappa}>1$ and $\overline{T}>1$, see Eq.~\eqref{eq:47}. In this case, the system is effectively in contact with a single PSN bath to which, by virtue of Eqs.~(\ref{eq:48}-\ref{eq:49}), the system dissipates heat $Q_a>0$ at an expense of  work $W_a>0$.

We explore the performance of the machine as a function of the PSN shot rate by exploring a wide range of orders of magnitude, from $\omega_a\tau_a\sim 10^{-3}$ to~$\omega_a\tau_a \sim 10^3$, by keeping all the rest of parameters fixed. To probe the non-Gaussian features of the particle statistics, we study here and in the following examples the  excess kurtosis  associated averaged over  the cycle, given by 
\begin{equation}
    \mathcal{K}_{\rm ex}^\mu = (   \displaystyle\mathcal{K}_{\rm ex}^{1}+ \displaystyle\mathcal{K}_{\rm ex}^2 )/2,
    \label{eq:kmean}
\end{equation}
\begin{figure*}[!htbp]
\centering
\includegraphics[width=0.8\textwidth]{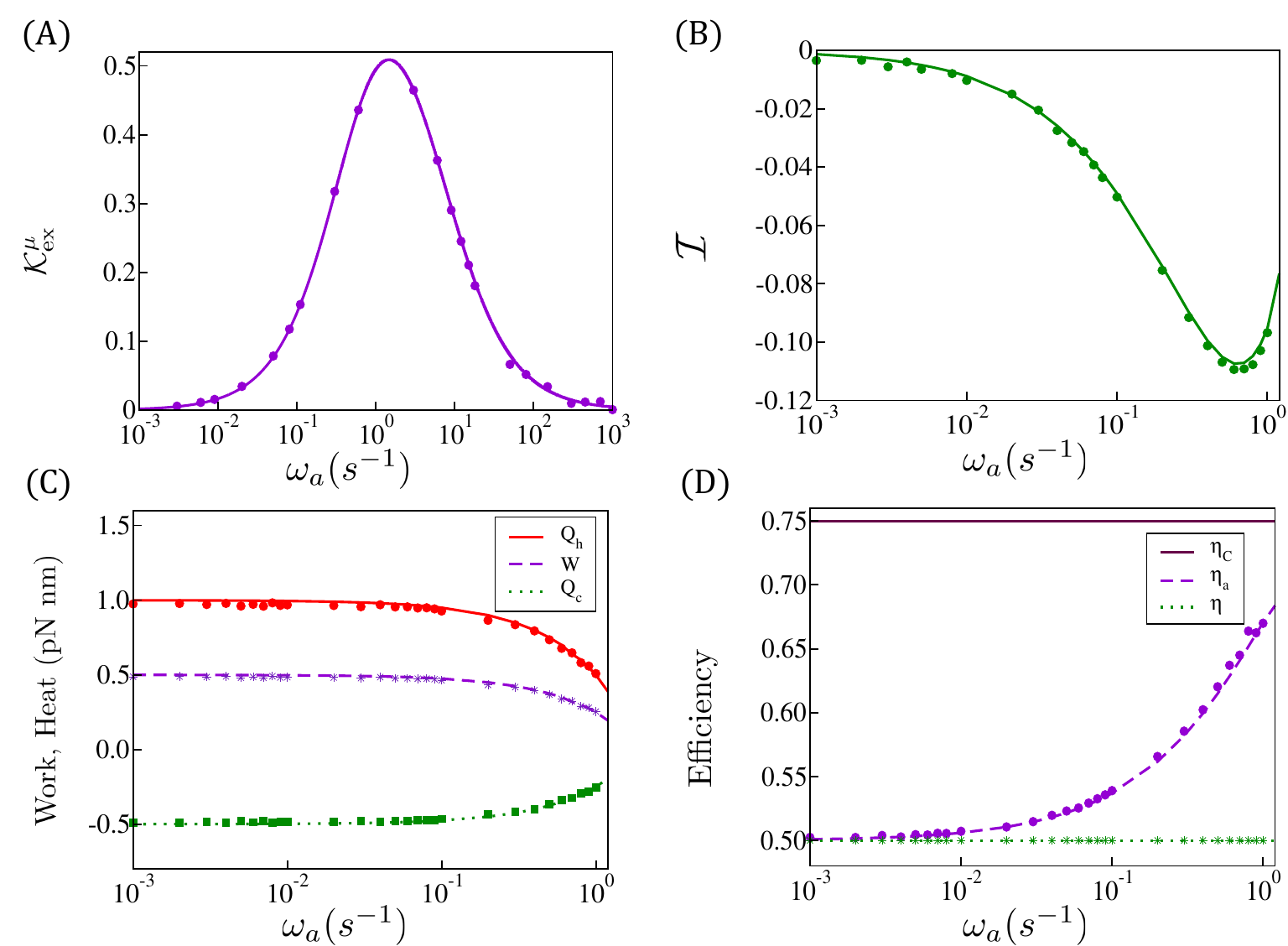}
\caption{Key dynamic and thermodynamic quantities associated with the two-stroke active heat engine  subject to thermal (Gaussian white noise) and active (Poisson-shot noise,   PSN) baths operating in the quasistatic limit with time-independent PSN shot rate, see first row in Table~\ref{TableRita}  and further details in Sec.~\ref{sec:Two_stroke_engine2}: analytical results (lines) and results from numerical simulations (symbols).  In all panels, we focus on the dependency of key dynamic/thermodynamic quantity as a function of the PSN shot rate $\omega_a$ while all other parameters are fixed to prescribed values (see below).
(A) Excess kurtosis $\mathcal{K}_{\rm ex}^\mu $ associated with the  position distribution averaged over the two strokes of the cycle, see Eq.~\eqref{eq:kmean}.  (B) Quasistatic divergence $\mathcal{I}$  defined by Eq.~\eqref{eq:I_inf} and given by  Eq. \eqref{I_general_psn} for this specific engine. (C) Total work extracted  $W$ (simulations in purple stars, theory in purple dashed line), total heat uptake from the hot baths $Q_h$ (simulations in red circles, theory in solid red line), and total heat dissipated to the cold baths $Q_c$ (simulations in green squares, theory in green dotted line), see Sec.~\ref{sec:Va}  for further details. (D) `Traditional' efficiency $\eta=W/Q_h$ (simulations in green circles, theory given by Eq.~\eqref{eq:pseudoeffi} in green dotted line) and `active' efficiency $\eta_a$ (simulations in purple circles, theory given by Eq.~\eqref{eta_I2} in purple dashed line). The maroon line is set to Carnot's efficiency $\eta_C=1-(T_c/T_h)$ in terms of the temperatures of the thermal baths. Other parameters are $\tau_a=2.0\,\mathrm{s}$, $T_{ac}=144.9\,\mathrm{K}$, $T_{ah}=144.9\,\mathrm{K}$, $T_c=72.5\,\mathrm{K}$, $T_h=289.9\,\mathrm{K}$ $\kappa_1=2.0\,\mathrm{pN/nm}$, $\kappa_2=4.0\,\mathrm{pN/nm}$, $\gamma=2.0\,\mathrm{pN\,s/nm}$, cycle time $\tau=100\,\mathrm{s}$, number of simulations  $10^5$ and simulation time step $dt=0.1\,\mathrm{ms}$.}
\label{FIG4}
\end{figure*}

where $\mathcal{K}_{\rm ex}^{1}$ and $\mathcal{K}_{\rm ex}^{2}$ are respectively the excess kurtosis of the accompanying distribution associated with parameter values $\Lambda_1$ and $\Lambda_2$, see Eq.~\eqref{eq:Kex} for their corresponding analytical expressions. Our numerical simulations, in excellent agreement with the analytical predictions, show that the mean excess kurtosis develops a non-monotonic behaviour with respect to the PSN shot rate (Fig.~\ref{FIG4}.~(A)): it is zero for $\omega_a\tau_a$ small where the active noise is negligible (equilibrium Gaussian fluctuations), and for $\omega_a\tau_a$ large where the active noise dominates yet approaching a Gaussian white noise limit at an effective temperature. For intermediate ranges, $\mathcal{K}_{\rm ex}^\mu \geq 0$ is positive, indicating a leptokurtic shape with fat tails. This exploration encouraged us to identify regions where the Gaussian approximation does not hold, leading us to investigate the engine behavior within the non-Gaussian regime.

To further dig into the non-equilibrium statistics of the engine, we evaluate the mean quasistatic divergence   $\mathcal{I}$ defined by  Eq.~\eqref{eq:I_inf} both analytically via Eq.~\eqref{eqn:ps_noneqsteady}  and with numerical simulations, within the PSN shot noise range where the system works as an engine (Fig.~\ref{FIG4}.~(B)). Interestingly, for this parameter choice we get $\mathcal{I}\leq 0$ negative, as a result of the leptokurtic shape of the accompanying distributions $\mathcal{K}_{\rm ex}^\mu \geq 0$. Notably, $\mathcal{I}$ inherits the non-monotonic behavior of the mean excess kurtosis, as both tend to zero in the small and large $\omega_a\tau_a$ limit, corresponding respectively to the equilibrium and the non-equilibrium Gaussian limit.
\begin{figure*}[!htbp]
\centering
\includegraphics[width=0.8\textwidth]{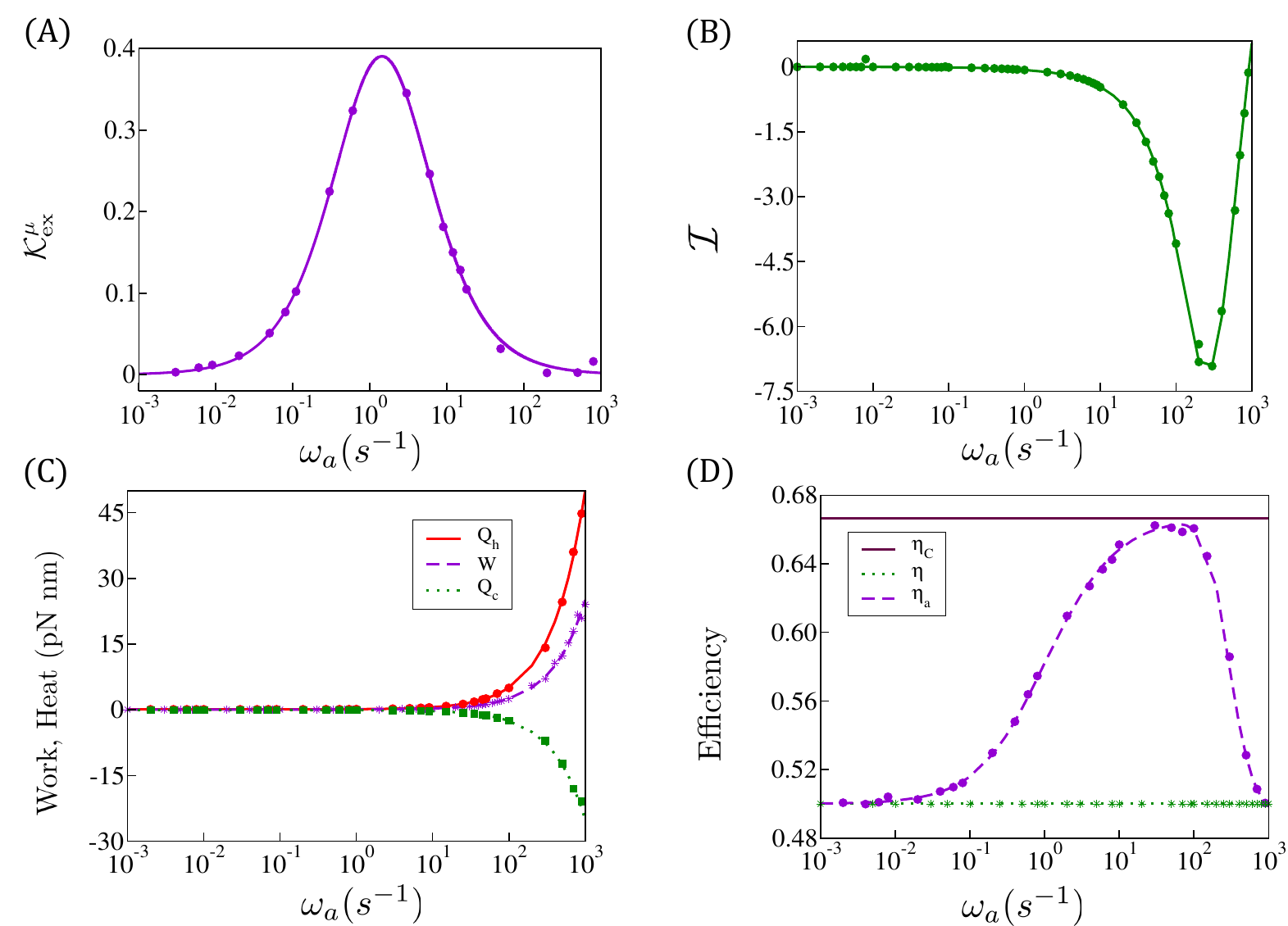}
\caption{Key dynamic and thermodynamic quantities associated with the two-stroke active heat engine  subject to thermal (Gaussian white noise) and active (Poisson-shot noise,   PSN) baths operating in the quasistatic limit with stepwise time-dependent  PSN shot rate, see second row in Table~\ref{TableRita}  and further details in Sec.~\ref{sec:Two_stroke_engine2}: analytical results (lines) and results from numerical simulations (symbols).  In all panels, we focus on the dependency of key dynamic/thermodynamic quantity as a function of the PSN shot rate $\omega_a$ while all other parameters are fixed to prescribed values (see below).
(A) Excess kurtosis $\mathcal{K}_{\rm ex}^\mu $ associated with the  position distribution averaged over the two strokes of the cycle, see Eq.~\eqref{eq:kmean}.  (B) Quasistatic divergence  $\mathcal{I}$  defined by Eq.~\eqref{eq:I_inf} and given by  Eq. \eqref{I_general_psn} for this specific engine. (C) Total work extracted  $W$ (simulations in purple stars, theory in purple dashed line), total heat uptake from the hot baths $Q_h$ (simulations in red circles, theory in solid red line), and total heat dissipated to the cold baths $Q_c$ (simulations in green squares, theory in green dotted line), see Sec.~\ref{sec:Vb}  for further details. (D) `Traditional' efficiency $\eta=W/Q_h$ (simulations in green circles, theory given by Eq.~\eqref{eq:pseudoeffi} in green dotted line) and `active' efficiency $\eta_a$ (simulations in purple circles, theory given by Eq.~\eqref{eta_I2} in purple dashed line). The maroon line is set to Carnot's efficiency $\eta_C=1-(T_c/T_h)$ in terms of the temperatures of the thermal baths. Parameters used are, $\tau_a=1.0\,\mathrm{s}$, $T_{ac}=21.7\,\mathrm{K}$, $T_{ah}=58.0\,\mathrm{K}$, $T_c=14.5\,\mathrm{K}$, $T_h=43.5\,\mathrm{K}$ $\kappa_1=0.1\,\mathrm{pN/nm}$, $\kappa_2=0.2\,\mathrm{pN/nm}$ and $\gamma=0.2\,\mathrm{pN\,s/nm}$,  cycle time $\tau=100\,\mathrm{s}$, number of simulations  $10^5$ and simulation time step $dt=0.1\,\mathrm{ms}$.}
\label{FIG5}
\end{figure*}

Next, we proceed with the non-equilibrium thermodynamic analysis by evaluating, both analytically and numerically, the  work and heat exchanges averaged over a cycle (Fig.~\ref{FIG4}.~(C)). For our parameter choice, the machine works as a two-stroke engine  for PSN shot rates exhibiting non-Gaussian behavior, up to  $\omega_a\tau_a \sim 20$, which is near the short rate at which the maximal value of the average kurtosis is attained (Fig.~\ref{FIG4}.~(A)).  Notably,  both the total heat uptake $Q_h$ and the  net total work extracted $W$ by the machine  decrease when increasing the PSN shot rate. However, their ratio (standard efficiency) $\eta=W/Q_h=1-(\kappa_1/\kappa_2)$ is independent of the PSN shot rate (see   green dotted line in Fig.~\ref{FIG4}.~(D)) as predicted by Eq.~\eqref{eq:pseudoeffi}. To probe the impact of activity in the efficiency it is thus imperative to adopt the active efficiency~$\eta_a$ from Eq.~\eqref{eta_I2} as a quantifier. Notably, $\eta_a$ captures the sensitivity on the PSN rate $\omega_a$. We find $\eta_a\geq \eta$ within the engine regime as a result of the leptokurtic statistics  which result in $\mathcal{I}\leq 0$. Furthermore, we find that $\eta_a$ increases with $\omega_a$ and approaches the Carnot limit at zero power. Therefore, under homogeneous activity $T_{\rm ah}=T_{\rm ac}$ the performance of the two-stroke   machine  at fast shot rates that resembles that of a  Carnot heat engine in the absence of active noise. See also in Appendix~\ref{app:two_step_protocol} and   Fig.~\ref{Fig:__plus_work_engine_regime_activity_overall}.~(A) for supporting additional data.

\subsection{\texorpdfstring{Heterogeneous activity ($\bar{\kappa}>1$, $\overline{T}>1$, $\overline{T_a}> 1$): thermal and active engines}{text}}
\label{sec:Vb}

As a second case study, we focus  in the   scenario in which the two (active) PSN baths are not identical, i.e. $\overline{T_a}>1$, $\overline{\kappa}>1$ and $\overline{T}>1$, see Eq.~\eqref{eq:47}. 
This scenario is plausible in e.g.  heat engines operating in  bacterial solutions~\cite{Krishnamurty16}, where changing the temperature of the solution may lead to changes in the bacterial activity (reproduction rate, self-propulsion velocity) that ultimately may impact the active noise intensities $T_{ac}$ and $T_{ah}$. 
In this case, the system is always in contact with a thermal and a PSN baths at temperatures $T_c$ and $T_{ac}$ in one half cycle and $T_h>T_c$ and $T_{ah}>T_h$ in the other half. As a result, the machine operates as two interacting heat engines, one passive with $Q_{\rm  th}>0$, $W_{\rm th}>0$ and another active with $Q_{\rm a }>0$, $W_{\rm a}>0$. 

As done in the previous case study in Sec.~\ref{sec:Va}, we also  evaluate in Fig.~\ref{FIG5},   analytically and numerically, key dynamical and thermodynamical quantities focusing on their dependency on the PSN shot rate $\omega_a$ for several orders of magnitude while keeping the rest of the parameters fixed. The mean excess kurtosis (Eq.~\eqref{eq:kmean}) displays also positive values (leptokurtic shape) following a non-monotonic behaviour between the small $\omega_a\tau_a$ and large $\omega_a\tau_a$ limits  corresponding to Gaussian statistics (Fig.~\ref{FIG5}.~(A)). 
Interestingly, here the quasistatic divergence $\mathcal{I}$ is also negative as a probe of leptokurtic behaviour with fat tails (Fig.~\ref{FIG5}.~(B)). However, for this parameter choice, $\mathcal{I}$  does not attain its maximum for $\omega_a$ values near  the maximal kurtosis but at several orders of magnitude larger. We attribute this phenomenon to the fact that at moderately large $\omega_a\tau_a$ the PSN amplitude is large yet being moderately non-Gaussian (see Eq.~\eqref{eq:LEPSN}). 
At odds with the time-homogeneous activity scenario, neither the heat uptake nor the work extracted per cycle tend to zero at $\omega_a\tau_a$ large (Fig.~\ref{FIG5}.~(C), cf. Fig.~\ref{FIG4}.~(C)). Indeed,  the two-stroke machine does not approach the `reversible' (zero power) limit at large $\omega_a$ since in this case the machine behaves as effectively being in contact with four thermal baths with anisotropic temperatures $\overline{T_a}\neq \overline{T}$. 
The leptokurtic shape inducing $\mathcal{I}$ negative is such that, by virtue of Eq.~\eqref{eta_I2} the active efficiency $\eta_a\geq \eta$ exceeds the traditional ($\omega_a$ independent) efficiency (Fig.~\ref{FIG4}.~(D)). 
Notably, in this parameter range the active efficiency $\eta_a$ attains its maximum at a value close to the Carnot limit for $\omega_a$ near the extremum of $\mathcal{I}$. This  result opens up the quest for Carnot efficiency at finite (non-zero) power~\cite{liang} to active systems.

\subsection{\texorpdfstring{Heterogeneous activity ($\bar{\kappa}>1$, $\overline{T}>1$, $\overline{T_a}> 1$): thermal refrigerator, active engine}{text}}

\label{sec:Vc}

So far in Sec.~\ref{sec:Va} and Sec.~\ref{sec:Vb} we analyzed parameter values with  $\overline{T}>\bar{\kappa}$ (i.e. $T_h/T_c > \kappa_2/\kappa_1 $) which ensure that the active baths operate as a heat engine. Here, we explore parameter values for which $\overline{T}\leq \bar{\kappa}$, where the thermal baths act as refrigerators while maintaining an overall engine behaviour sustained by the PSN baths operating as heat engines. For all parameter values, we work in the overall heat engine operating regime in which the total work per cycle is extracted $W>0$ at the expense of total input heat $Q_h>0$.

\begin{figure}[htpb!]
\hspace{-0.5cm}
\includegraphics[width=0.45\textwidth]{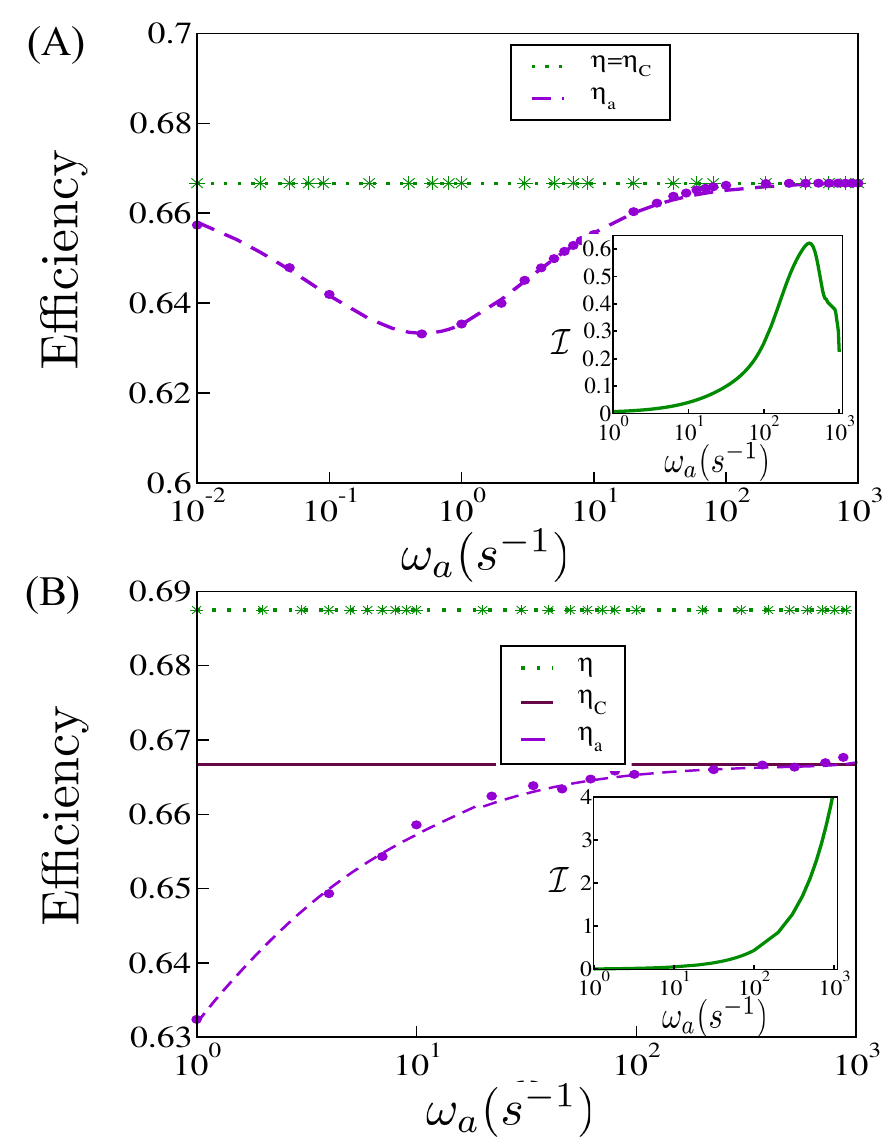}\hfill
\caption{Efficiency associated with the two-stroke active heat engine  subject to thermal (Gaussian white noise) and active (Poisson-shot noise,   PSN) baths operating in the quasistatic limit with stepwise time-dependent  PSN shot rate, see third row in Table~\ref{TableRita}  and further details in Sec.~\ref{sec:Vc}: analytical results (lines) and results from numerical simulations (symbols).  In all panels, we focus on the dependency of key dynamic/thermodynamic quantity as a function of the PSN shot rate $\omega_a$ while all other parameters are fixed to prescribed values (see below).
`Traditional' efficiency $\eta=W/Q_h$ (simulations in green circles, theory given by Eq.~\eqref{eq:pseudoeffi} in green dotted line) and `active' efficiency $\eta_a$ (simulations in purple circles, theory given by Eq.~\eqref{eta_I2} in purple dashed line). The maroon line is set to Carnot's efficiency $\eta_C=1-(T_c/T_h)$ in terms of the temperatures of the thermal baths. The two panels correspond to two different parameter choices: (A) $\kappa_2=0.3\,\mathrm{pN/nm}$ for which   $\kappa_1/\kappa_2=T_c/T_h$; and (B)  $\kappa_2=0.32\,\mathrm{pN/nm}$ for which $\kappa_1 / \kappa_2 <T_c / T_h$.  The rest of the parameters were set equal for (A) and (B) and given by $\tau_a=1.0\,\mathrm{s}$, $T_{ac}=14.5\,\mathrm{K}$, $T_{ah}=58.0\,\mathrm{K}$, $T_c=14.5\,\mathrm{K}$, $T_h=43.5\,\mathrm{K}$, $\gamma=0.2\,\mathrm{pN\,s/nm}$, $\kappa_1=0.1\,\mathrm{pN/nm}$,  cycle time $\tau=100\,\mathrm{s}$, number of simulations  $10^5$ and simulation time step $dt=0.1\,\mathrm{ms}$.}
\label{Fig:effi_etaeqletac}
\end{figure}

First, we study the limit case $\bar{T}=\bar{\kappa}$, i.e. for $\kappa_1 /\kappa_2 = T_c /T_h$. In this case, the `traditional' efficiency $\eta=W/Q_h$ equals to the Carnot limit $\eta_C=1-(T_c/T_h)$ associated with the temperatures of the thermal baths, which is achieved at vanishing $W_{\rm th}$ and $Q_{\rm th}$, see Eqs.~(\ref{eq:48}) and~(\ref{eq:49}). Concomitantly,  $W_{\rm a}$ and $Q_{\rm a}$, are positive (active heat engine behavior) when ensuring an overall engine operation. Fig.~\ref{Fig:effi_etaeqletac}.~(A) shows that the active efficiency $\eta_a\leq \eta_C = \eta$ is, for all $\omega_a$, smaller than the traditional efficiency, which is in stark contrast with the opposite behavior ($\eta_a\geq  \eta$) found in for $\bar{T}>\bar{\kappa}$ in Sec.~\ref{sec:Va} and Sec.~\ref{sec:Vb}.  The definition of $\eta_a$ given by Eq.~\eqref{eta_I2} implies that  the quasistatic divergence $\mathcal{I}$ should be positive such that $\eta_a\leq \eta$ in the engine regime, a result that we verify with our numerical simulations in the inset of   Fig.~\ref{Fig:effi_etaeqletac}.~(A).

Even more striking is the case where $\bar{T}<\bar{\kappa}$, i.e. when $\kappa_1 /\kappa_2 >T_c /T_h$, which, by virtue of Eq.~\eqref{eq:pseudoeffi} leads to apparent Super Carnot 'traditional' efficiencies $\eta>\eta_C$, as confirmed by our numerical simulations in Fig.~\ref{Fig:effi_etaeqletac}.~(B). As for the limit case $\bar{T}= \bar{\kappa}$, here the quasistatic divergence~$\mathcal{I}$ is also positive (see inset of Fig.~\ref{Fig:effi_etaeqletac}.~(B)), thus being a statistical footprint of an active efficiency $\eta_a$ lying below the traditional value $\eta$. 

\subsection{Relation between active efficiency and non-Gaussianity}
To further investigate the relationship between the engine performance and the degree of non-Gaussianity of the particle statistics,  Fig.~\ref{Fig:effi_exkurt_PC}. shows, the active efficiency $\eta_a$ as a function of the mean excess kurtosis $K^{\mu}_{\rm ex}$. The color of the data points indicates the corresponding PSN shot rate $\omega_a$. For the homogeneous activity case in Fig. \ref{Fig:effi_exkurt_PC}.~(A),~$\eta_a$ increases monotonously with $K^{\mu}_{\rm ex}$. This result implies that, in this parameter regime, a larger efficiency is associated with a larger non-Gaussianity. However, the relation is not linear. This is because $K^{\mu}_{\rm ex}$ describes only one property of the position distribution, while $\eta_a$ depends on the quasistatic divergence $\mathcal{I}$, which contains information about the full difference between the accompanying and equilibrium distributions. A different behavior emerges for the  case with time-dependent activity in, see Fig.~\ref{Fig:effi_exkurt_PC}.~(B) where the relation between $\eta_a$ and $K^\mu_{\rm ex}$ is non-bijective.  This occurs because $K^\mu_{\rm ex}$ varies non-monotonically with $\omega_a$, while the
active efficiency follows a different dependence on the activity. Consequently, the excess kurtosis alone is not sufficient to determine the thermodynamic performance of the engine. 
Excess kurtosis has also been used to characterize non-Gaussian heat fluctuations in passive and active overdamped systems, where kinetic-energy changes can play an important role in the heat statistics \cite{Paraguassu2025}. In our work, however, we consider the excess kurtosis of the particle position distribution and investigate its relation to the active efficiency.

\section{Discussion and Outlook}
\label{sec:Discussion_final}

\begin{figure}[!t]
\includegraphics[width=0.475\textwidth]{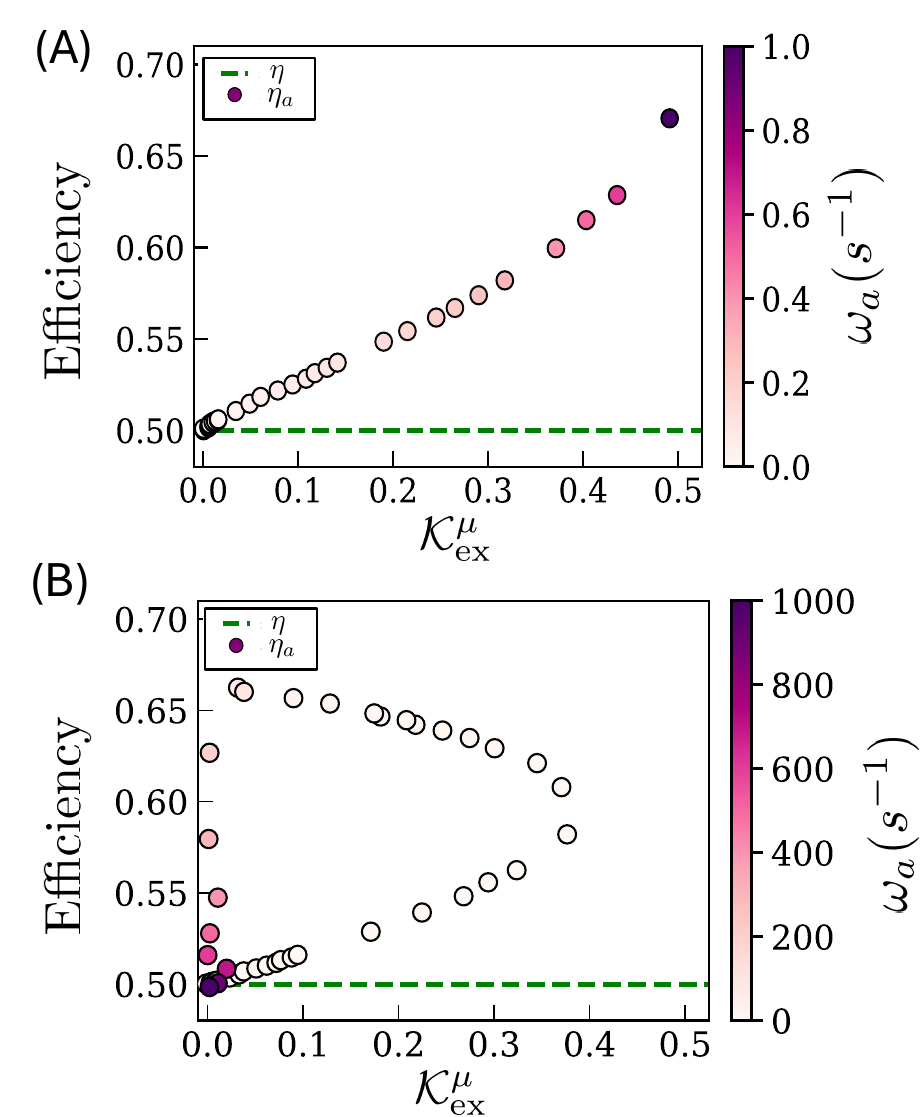}\hfill
\caption{Active efficiency $\eta_a$ (Eq.~\eqref{eta_I2}) as a function of the mean excess kurtosis $\mathcal{K}_{\rm ex}^\mu$ (Eq.~\eqref{eq:kmean}) associated with the quasistatically-driven two-stroke active heat engine. Panels (A) and (B) show the data corresponding to Fig.~\ref{FIG4} and  \ref{FIG5} respectively.
In both panels, the horizontal green line is the traditional efficiency $\eta$ (Eq.~\eqref{eq:pseudoeffi}). The Carnot efficiencies associated with the thermal
baths are $\eta_C=1-T_c/T_h\simeq 0.75$ for
panel (A) and $\eta_C\simeq 0.667$ for panel (B). The color bar illustrates the value of $\omega_a$ associated with each point of the scatter plot.}
\label{Fig:effi_exkurt_PC}
\end{figure}

In this work we have put forward with key examples recent theory~\cite{Barato,Baratoref26} regarding the thermodynamic performance of stochastic heat engines and refrigerators operating under generic non-conservative (e.g. active) forces. To this aim, we focused on two-stroke and Stirling-like protocols executed on small machines described by Langevin equations that are in contact with two thermal baths and two athermal baths described by Poissonian shot noise~(PSN)~\cite{Edgar24}.  Such minimalistic stochastic models reveal intriguing  phenomena which allow, for instance, simultaneous engine and refrigerator regimes with respect to the thermal and/or athermal environments. Here, we have focused, without loss of generality, on the quasistatic limit operation for parameter values in which the overall performance is that of an engine. Our quantification of thermodynamic and thermodynamic quantities revealed an intriguing interplay between non-Gaussian statistics, efficiency, and the degree of non-equilibrium quantified by  a purely statistical quantity, the quasistatic divergence, for which we give refreshing intuitive insights in our example model.

A key insight of our work is that relying in traditional definitions of efficiency (work extracted per heat intake) must be taken with a grain of salt, especially in the presence of non-equilibrium (e.g. active, PSN) forces. 
Indeed, for some parameter values the traditional efficiency  exceeds  the Carnot limit (with respect to the thermal baths' temperatures) even in the quasistatic limit. In such situations the so-called active efficiency, which takes into account the excess entropy generated by the active processes, remains a reliable thermodynamically-consistent quantity always limited by the Carnot bound. 
Such active efficiency is directly measurable in experiments of e.g. colloidal heat engines in bacterial or PSN baths~\cite{Krishnamurty16,Krishnamurty21} as it is expressed in terms of total heat and work averages and the quasistatic divergence which is a purely statistical quantity that does not require prior knowledge on any physical parameter value.

Taken together, our results not only rationalize efficiency in the presence of active noise but also provide a link between thermodynamics and geometry of the quasistatic distribution. Experimentalists may access directly to the quasistatic divergence from samples of the (generally non-Gaussian) distribution of the particle position, and use it to describe the machines' efficiency without needing to rely on crude equilibrium approximations using effective temperatures~\cite{roy2025single}.  
In addition, the notion of a bonafide `active' efficiency sheds light on the relation between performance and non-Gaussianity as a probe of non-Boltzmannian statistics.  

Experimental work~\cite{Krishnamurty16} suggested boosts in bacterial heat engine's efficiency as due to non-Gaussian statistics, yet focusing on quantifications with effective temperatures.
Fig.~\ref{Fig:effi_exkurt_PC}. reveals that,  the bonafide `active' efficiency may or may not increase with the degree of non-Gaussianity (average excess kurtosis).

In our proof-of-principle simulations, we find that for the case of homogeneous PSN activity throughout the cycle, the active efficiency correlates with the excess kurtosis through a non-linear relationship (Fig.~\ref{Fig:effi_exkurt_PC}.~(A)). On the other hand, for non-homogeneous active environments (as those expected for say bacterial baths) the active efficiency  and kurtosis do not correlate strongly, developing  a complex non-bijective relation between them (Fig.~\ref{Fig:effi_exkurt_PC}.~(B)). In particular, the results of Fig.~\ref{Fig:effi_exkurt_PC} show that the excess kurtosis alone does not generally determine the active efficiency.  Therefore, our results imply that characterizing the departure from equilibrium through a single measure of non-Gaussianity is often insufficient to fully describe  the underlying thermodynamic performance. The quasistatic divergence $\mathcal{I}$ provides a more detailed description in this respect, since it depends on the full accompanying and equilibrium distributions. This also makes $\mathcal{I}$ useful from an experimental point of view, as it can in principle be estimated from the measured position distributions without requiring detailed knowledge of the microscopic origin of the active fluctuations. Altogether, our results rationalize pioneering experimental results tackling the stochastic thermodynamics of small machines in the presence of generic non-equilibrium reservoirs. 

To reinforce and generalize our conclusions, we also explored in the Appendices additional scenarios involving variations in active temperatures, spatial confinement, and temporal distribution of activity. These results collectively emphasize that active noise sources may enhance engines' performance, but the degree of enhancement is highly dependent on how activity shapes the distance to equilibrium statistics throughout the thermodynamic cycle. In summary, our findings reveal a rich interplay between active noise, information, and thermodynamic performance. 

\section*{Acknowledgements}
We thank Gonzalo Manzano (IFISC) for initial discussions on the formulation of the model. RiM acknowledges funding from the ICTP STEP (Sandwich Training Educational Program). RalM acknowledges funding from the German Research Foundation (DFB, grant No. 318763901, CRC Data Assimilation).

\bibliography{reference}
\newpage
\onecolumngrid
\appendix
\counterwithin{figure}{section}
\counterwithin{table}{section}
\renewcommand\thefigure{\thesection\arabic{figure}}    
\section*{Appendix}
In the Appendices below we give additional details and supporting material on the results shown in the Main Text. The Appendices are arranged as follows. In Appendix \ref{app:FPE_sol}, we derive the solution of the Fokker-Planck equation namely Eq. (\ref{eqn:FokkerPlanck_solution}). In Appendix \ref{app:fokker_planck}, we calculate the moments and the excess kurtosis for constant trap strength $\kappa$. Appendices \ref{app:two_step_protocol} and \ref{app:twostroke_add} are dedicated to the calculations of mean squared displacements and different thermodynamic quantities for the two-stroke engine protocol discussed in Sec. \ref{sec:Two_stroke_engine}. In Appendix \ref{app:stirling_protocol} we repeat the calculations for the Stirling protocol. Appendix \ref{sec:code} deals with the details of the numerical simulation techniques used in the manuscript.  
\section{Fokker-Planck equation solution}
\label{app:FPE_sol}
Here we provided the detailed derivation of formula~\eqref{eqn:FokkerPlanck_solution} in the Main Text.
Our derivation begins by considering the Fourier transform of the propagator, defined as
\begin{equation}
    \hat{P}(q,t) \equiv \int_{-\infty}^{\infty} e^{iqx} P(x,t) dx.
\end{equation}
After Fourier transforming the Fokker-Planck Equation~\eqref{eqn:FokkerPlanck}, we get
\begin{equation}
    \dfrac{\partial \hat{P}(q,t)}{\partial t} = -\dfrac{\kappa(t)}{\gamma} q \dfrac{\partial \hat{P}(q,t)}{\partial q} - \dfrac{k_B T}{\gamma} q^2 \hat{P}(q,t) + \omega_a \hat{P}(q,t) (\hat{\rho}_a(q,t) - 1),\label{eq:A2}
\end{equation}
with vanishing probability for $|q|\to \infty$ and with $\hat{P}(q,t_0)= e^{iqx_0}$ the Fourier transform of the initial distribution $P(x,t_0)=\delta(x-x_0)$. In Eq.~\eqref{eq:A2}, we used the convolution property for the Fourier transform at the right hand side. Equation~\eqref{eq:A2} can be rewritten as
\begin{equation}
    \dfrac{\partial (\ln \hat{P}(q,t))}{\partial t} = -\dfrac{\kappa(t) }{\gamma}q \dfrac{\partial (\ln \hat{P}(q,t))}{\partial q} - \dfrac{k_BT}{\gamma} q^2 + \omega_a (\hat{\rho}_a(q,t)-1).\label{eq:A3}
\end{equation}
Equation~\eqref{eq:A3} is a hyperbolic partial differential equation (PDE) in the function $\ln \hat{P}(q,t)$ which can be solved through the method of characteristics. For simplicity let us define the function $R(q,t) \equiv \ln \hat{P}(q,t)$, which allow us to rewrite the PDE, Eq.~\eqref{eq:A3} as
\begin{equation}
    \label{eqn:R_PDE}
    \dfrac{\partial R(q,t)}{\partial t} + \dfrac{\kappa(t)}{\gamma} q \dfrac{\partial R(q,t)}{\partial q} = -\dfrac{k_BT}{\gamma}q^2 + \omega_a(\hat{\rho}_a(q,t) -1).
\end{equation}
We now solve Eq.~\eqref{eqn:R_PDE} by making the following ansatz. We suppose that the variable $q$ is actually a function of another variable $s$ (characteristic curves). As a result, the total derivative of $R(q(s),s)$ with respect of $s$ reads
\begin{equation}
    \dfrac{d R(q(s),s)}{d s} = \dfrac{dq}{ds}\dfrac{\partial R(q(s),s)}{\partial q} + \dfrac{\partial R(q(s),s)}{\partial s}.
\end{equation}
The trick to simplify, Eq.~\eqref{eqn:R_PDE} is to set
\begin{equation}
    \dfrac{dq}{ds} = \dfrac{\kappa(s)}{\gamma} q 
\end{equation}
with the final condition $q(t)=q$, which gives
\begin{equation}
    \label{eqn:q(s)}
    q(s) = \exp \left( \int_{s}^{t} \dfrac{\kappa(s')}{\gamma} ds' \right) q = \varphi(t,s) q
\end{equation}
where we set
\begin{equation}
    \varphi(t,s) \equiv \exp \left(\int_{s}^t \dfrac{\kappa(s')}{\gamma} ds' \right).
\end{equation}
Having this position on the variable $q$, the PDE, Eq.~\eqref{eqn:R_PDE} reduces to 
\begin{equation}
    \dfrac{d R(q(s),s)}{ds} = -\dfrac{k_BT}{\gamma}q(s)^2 + \omega_a(\hat{\rho}_a(q(s),s)-1)
\end{equation}
which is a first-order ordinary differential equation (ODE), its solution reads
\begin{equation}
    R(q(s),s) = R(q(t_0),t_0) -\dfrac{k_BT}{\gamma} \int_{t_0}^{s} q(s')^2 ds' +\omega_a\int_{t_0}^{s} (\hat{\rho}_a(q(s'), s') -1) ds',\label{eq:A11}
\end{equation}
by using the initial condition on $R(q(t_0),t_0)= i q(t_0) x_0= i qx_0 \varphi(t_0,s)$, and the explicit form of $q(s)$ we get
\begin{equation}
    R(q(s),s) = iq x_0 \varphi(t_0,s) - \dfrac{k_BT}{\gamma} q^2 \int_{t_0}^{s} \varphi(t,s')^2  ds' + \omega_a \int_{t_0}^{s} \left[\hat \rho_a \left( \varphi(t,s') q,s' \right) -1 \right] ds',
\end{equation}
evaluating the last expression for $s=t$ and reminding that $R(q,t) = \ln \hat P (q,t)$ we finally get
\begin{equation}
    \ln \hat P(q,t) = iq \left( x_0 \varphi(t,t_0)^{-1} \right) - \dfrac{k_BT}{\gamma} q^2 \dfrac{\int_{t_0}^{t} \varphi(s,t_0)^2 ds}{\varphi(t,t_0)^2}  + \omega_a \int_{t_0}^{t} \left[\hat \rho_a \left( q \dfrac{ \varphi(s,t_0)}{\varphi(t,t_0)},s \right) -1 \right] ds,
\end{equation}
where we used the fact that $\varphi(t,s) = \varphi(s,t_0) / \varphi(t,t_0)$ and that $\varphi(t_0,t) = \varphi(t,t_0)^{-1}$. After exponentiation and Fourier inversion of this equation one gets formula~\eqref{eqn:FokkerPlanck_solution}  in the Main Text. 

In the specific case in which the kicks' distribution $\rho_a$ does not depend on time, it is possible to rewrite the previous integral in a nicer form by making a change of variables 
\begin{equation}
    \ln \hat{P}(q,t) = iq \left( x_0 \varphi(t,t_0)^{-1}\right) -\dfrac{k_BT}{\gamma} q^2 \dfrac{\int_{t_0}^t \varphi(t',t_0)^2 dt'}{\varphi(t,t_0)^2} +
    \dfrac{\omega_a \gamma}{\kappa(t)} \int_{\frac{q}{\varphi(t,t_0)}}^{q} \dfrac{\hat{\rho}(z)-1 }{z} dz,
\end{equation}
from which the accompanying density in Eq.~\eqref{eqn:ps_noneqsteady} can be easily obtained by setting the parameters constant and taking $t\to \infty$.

\section{Variance, skewness and kurtosis of the accompanying density}
\label{app:fokker_planck}
In this Section, we report the expressions for the variance, the skewness and the excess kurtosis of the accompanying density $P^{\rm ac}$ of the stochastic process $x$. They are respectively defined as
\begin{equation}
\mathrm{Var}[x]\equiv\langle(x-\langle x\rangle)^2\rangle,\quad 
\Tilde{\mu}_3[x]\equiv\dfrac{\langle(x-\langle x\rangle)^3\rangle}{\mathrm{Var}
[x]^{3/2}},\quad\mathcal{K}_{\rm ex}[x]\equiv\frac{\langle (x-\langle x \rangle)
^4\rangle}{\mathrm{Var}[x]^{2}}-3,
\label{skew_kurt_formula}
\end{equation}
and their analytical expressions read,
For the accompanying state, the corresponding analytical expressions are
\begin{align}
\langle x \rangle &= \frac{\gamma \omega_a}{\kappa}\,\langle Y\rangle_a, \\
\mathrm{Var}[x] &= \frac{k_B T}{\kappa}+\frac{\gamma \omega_a}{2\kappa}\langle Y^2\rangle_a,
\label{eqn:x_mean_var_engine} \\
\tilde{\mu}_3[x] &= \sqrt{\frac{8}{9}}\,
\frac{\gamma \omega_a}{\kappa}\,
\langle Y^3\rangle_a
\left(
\frac{2k_B T}{\kappa}
+\frac{\gamma \omega_a}{\kappa}\langle Y^2\rangle_a
\right)^{-3/2}, \\
\mathcal{K}_{\rm ex}[x] &= 
\frac{\gamma \omega_a}{\kappa}\,
\langle Y^4\rangle_a
\left(
\frac{2k_B T}{\kappa}
+\frac{\gamma \omega_a}{\kappa}\langle Y^2\rangle_a
\right)^{-2}.
\label{eq:exsskurtosis_psn}
\end{align}

Since \(\rho_a(y,t)\) is a zero-mean Gaussian distribution, all odd moments vanish. Therefore,

\begin{equation}
\begin{aligned}
\langle Y\rangle_{a} &=0,
&\qquad
\langle Y^3\rangle_{a} &=0
\end{aligned}
\label{eq:C1_general}
\end{equation}
\begin{equation}
\begin{aligned}
\langle Y^2\rangle_{a}&= \frac{\tau_a k_{\rm B}T_a}{\gamma},~
&\qquad
\langle Y^4\rangle_{a} 
=3\left(
    \frac{\tau_a k_{\rm B}T_a}{\gamma}
    \right)^2 .
\end{aligned}
\label{eq:C2_general}
\end{equation}

The general expressions of these statistical quantities were derived in Ref.~\cite{Edgar24}. 

\section{Calculation for two-stroke engine protocol}
\label{app:two_step_protocol}
 In this section, we  derive $\langle x_{i}^2(t)\rangle$ for each stroke starting from the formal solution of the Langevin equation given by Eq. \eqref{eqn:Langevin_2}, while keeping the initial value of the stroke arbitrary. 
\begin{equation}
x_i(t)=\exp\left(-\frac{1}{\gamma}\int^t{dt^{\prime}\kappa_i(t^{\prime})}\right) \left[x_{0i} +\frac{1}{\gamma}
    \int^t {dt^{\prime}~\left(\sqrt{2k_BT(t')\gamma}\xi(t')+\gamma \zeta_a(t')\right)}\exp\left(\frac{1}{\gamma}\int^{t'}{dt''\kappa_i(t'')}\right) \right].
\label{formal}
\end{equation} 
 where $i\in{1,2}$ labels the two strokes of the cycle and $x_{0i}$ refers to the initial position. Since the kick distribution has zero mean in both parts of the cycle, the periodic mean position is zero, ($\langle x_i(t)\rangle=0$). During the first half of the cycle $0 \leq t <\tau/2$ the stiffness, temperature and PSN effective temperature are set to $\kappa_1$, $T_c$, and $T_{ac}$ respectively, we get,
\begin{equation}
\langle x^2_{1}(t)\rangle =  \langle x^2(0)\rangle \exp{\left(-\dfrac{2\kappa_1 t}{\gamma}\right)}+ F_{1}(t)~,~~~ F_{1}(t)=\left(\dfrac{k_{B}T_c}{\kappa_1}+\dfrac{\gamma \omega_a}{2\kappa_1}\langle Y^2\rangle_{ac}\right)\left(1-\exp{\left(-\dfrac{2\kappa_1 t}{\gamma}\right)}\right).
\label{eq:xsqr1}
\end{equation}
Within the second half of the cycle $\tau/2 \leq t <\tau$ during which the stiffness, temperature and PSN effective temperature are set to $\kappa_2$, $T_h$, and $T_{ah}$ respectively, we get,
\begin{equation}
\langle x^2_{2}(t)\rangle =  \langle x(\tau/2)^2\rangle \exp{\left(-\dfrac{2\kappa_2}{\gamma}(t-\tau/2)\right)}+ F_{2}(t)~,~~~F_{2}(t)=\left(\dfrac{k_{B}T_h}{\kappa_2}+\dfrac{\gamma \omega_a}{2\kappa_2}\langle Y^2\rangle_{ah}\right)\left(1-\exp{\left(-\dfrac{2\kappa_2 (t-\tau/2)}{\gamma}\right)}\right).
\label{eq:xsqr2}
\end{equation}
The functions $F_1(t)$ and $F_2(t)$ represent the noise-induced contributions accumulated during the first and second strokes, respectively. They contain both the passive thermal contribution and the active PSN contribution. 
The kick amplitudes are drawn from the zero-mean Gaussian distribution ($\rho_a(y,t)$) introduced in Eq.~\eqref{kickGauss}. During the first and second half-cycles, the corresponding active ensembles are characterized by the effective active temperatures $T_{ac}$ and $T_{ah}$, respectively. Hence, using Eqs.~\eqref{eq:6},~\eqref{eq:C2_general} and~\eqref{eq:C1_general} the first and second moments of the kick-amplitude distributions are
\begin{equation}
\begin{aligned}
\langle Y\rangle_{ac} &=0,
&\qquad
\langle Y^2\rangle_{ac} &=
\frac{\tau_a k_{\rm B}T_{ac}}{\gamma},
\end{aligned}
\label{eq:C1}
\end{equation}
\begin{equation}
\begin{aligned}
\langle Y\rangle_{ah}&=0,~
&\qquad
\langle Y^2\rangle_{ah} 
=\frac{\tau_a k_{\rm B}T_{ah}}{\gamma}.
\end{aligned}
\label{eq:C2}
\end{equation}

The subscripts ``ac'' and ``ah'' distinguish the statistics of the active kicks during the first and second half-cycles, respectively. Imposing continuity at \(t=\tau/2\), $\langle x^2(\tau/2^-)\rangle=\langle x^2(\tau/2^+)\rangle$ together with periodicity over a complete cycle $
\langle x^2(0)\rangle=\langle x^2(\tau)\rangle$, 
we obtain
\begin{eqnarray}
\label{eq:apn_x^2}
\langle x^2(0)\rangle=\dfrac{ F_1(\tau/2) \exp{\left(-\dfrac{\kappa_2\tau}{\gamma}\right)}+F_{2}(\tau)}{\left(1-\exp{\left(-\dfrac{(\kappa_1+\kappa_2)\tau}{\gamma}\right)}\right)},~~\langle x^2(\tau/2^-)\rangle=\langle x^2(0)\rangle \exp{\left(-\dfrac{\kappa_1\tau}{\gamma}\right)}+F_{1}(\tau/2).
\end{eqnarray}
 Therefore, using Eqs. \eqref{eq:C1} and \eqref{eq:C2}, we finally get,
\begin{equation}
\label{eqn:mwan_var_engine}
\langle x(t)\rangle=0,\quad\langle x^2(0)\rangle=\langle x^2(\tau)\rangle =\dfrac{k_BT_h}{\kappa_2}+
\dfrac{\omega_a\tau_a k_BT_{ah}}{2\kappa_2},
\end{equation}
and
\begin{equation}
\label{eqn:mwan_var_engine_2}
\quad\langle x^2(\tau/2^{+})\rangle=\langle x^2(\tau/2^{-})\rangle =\dfrac{k_BT_c}{\kappa_1}+
\dfrac{\omega_a\tau_a k_BT_{ac}}{2\kappa_1}.
\end{equation}
We now use the finite-time solution above for $(\langle x^2(t)\rangle)$ to analyze the thermodynamic performance of the two-stroke engine. In particular, we consider heterogeneous activity and focus on the dependence of the thermodynamic quantities on the cycle duration ($\tau$), while all other parameters are kept fixed.

Figure~\ref{Fig:effi_tau_work_engine_regime} compares the traditional and active efficiencies with the Carnot limit as a function of the cycle duration~$\tau$. 
When $\tau$ is small, the system does not have enough time to relax within each stroke.
In this regime, the quantities in Fig.~\ref{Fig:effi_tau_work_engine_regime} shows a noticeable dependence on $\tau$, in particular, the active efficiency $\eta_a$ displays its largest deviation from its long-$\tau$ value, and the combination $\mathcal{I}+\beta_c\eta_C Q_h$ remains comparatively small (Fig.~\ref{Fig:effi_tau_work_engine_regime}~(B)). 
As $\tau$ is increased, relaxation becomes more effective during each stroke and the cycle progressively resembles a quasistatic transformation, in the sense that the engine follows the accompanying state more closely. 
 $\eta_a$ becomes nearly $\tau$ independent and $\mathcal{I}+\beta_c\eta_C Q_h$ reaches a higher, roughly constant plateau. 
In the present data, this crossover from fast, non-quasistatic operation to slow, near-quasistatic behavior occurs around $\tau\simeq 5$, where $\mathcal{I}+\beta_c\eta_C Q_h$ exhibits a clear increase and the efficiencies subsequently stabilize.

\begin{figure*}[h]
\centering
\includegraphics[width=\textwidth]{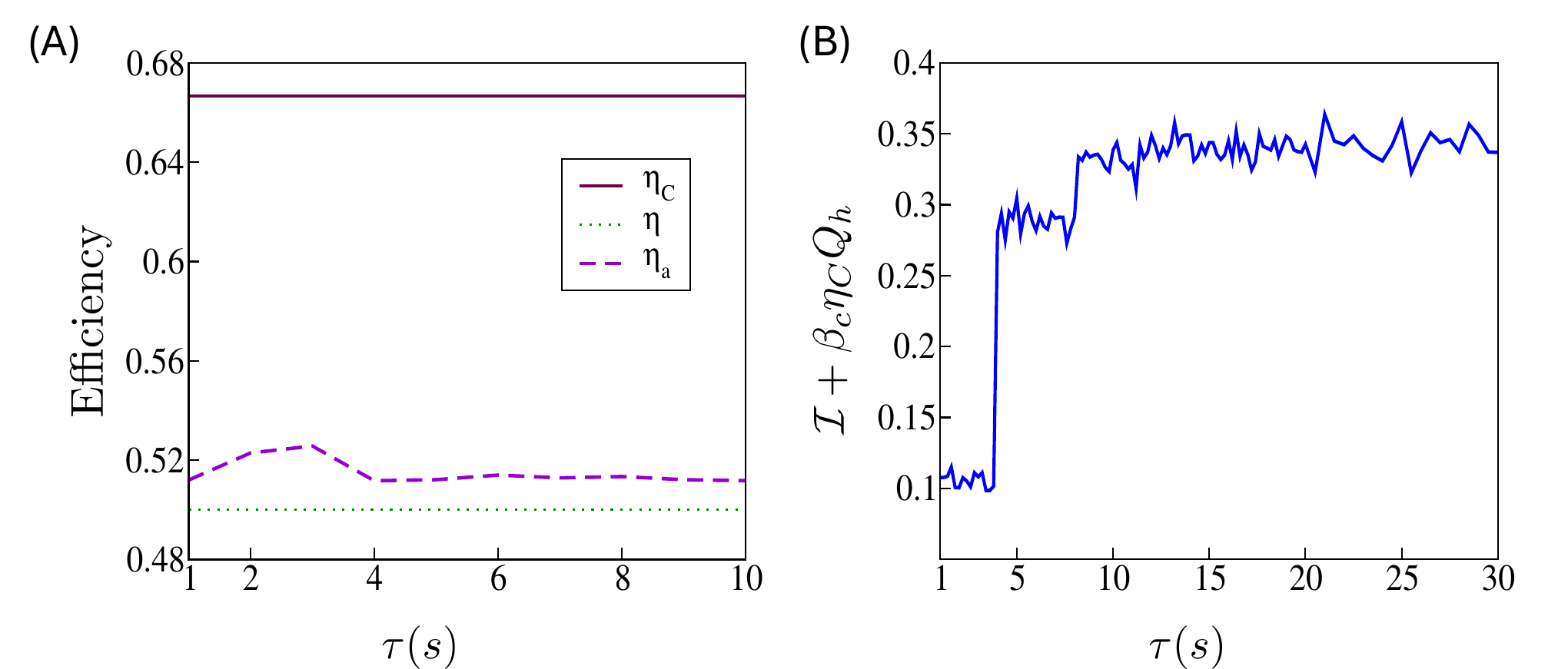}
\caption{Key thermodynamic quantities of the two-stroke active heat engine subject to thermal Gaussian white noise and active Poisson-shot noise baths. 
(A) Active efficiency \(\eta_a\) as a function of \(\tau\), showing the crossover from the non-quasistatic to the quasistatic regime. `Traditional' efficiency $\eta=W/Q_h$ (simulations in green dotted line) and `active' efficiency $\eta_a$ (simulations in purple dashed line). The maroon line is set to Carnot's efficiency $\eta_C=1-(T_c/T_h)$ in terms of the temperatures of the thermal baths
(B) The quantity \(\mathcal{I}+\beta_c \eta_C Q_h\) as a function of \(\tau\).
Parameters used are, $\tau_a=1.0\,\mathrm{s}$, $T_{ac}=14.5\,\mathrm{K}$, $T_{ah}=58.0\,\mathrm{K}$, $T_c=14.5\,\mathrm{K}$, $T_h=43.5\,\mathrm{K}$, $\kappa_1=0.1\,\mathrm{pN/nm}$, $\kappa_2=0.2\,\mathrm{pN/nm}$ and $\gamma=0.2\,\mathrm{pN\,s/nm}$, $\omega_a=0.1\,\mathrm{s}^{-1}$ and number of simulations  $10^5$ and simulation time step $dt=0.1\,\mathrm{ms}$.}
\label{Fig:effi_tau_work_engine_regime}
\end{figure*}

\section{Additional results for two-stroke engine}
\label{app:twostroke_add}
In this section, we present additional results related to the active-efficiency bound discussed in Sec.~\ref{stocengergy}. From the second law for the excess entropy production given by Eq.~\eqref{eq:datta}, the active efficiency defined in Eq.~\eqref{eta_I} is bounded by the Carnot efficiency provided that
\begin{equation}
\mathcal{I}+\beta_c\eta_C Q_h \geq 0.
\end{equation}
\begin{figure*}[!htbp]
\centering
\includegraphics[width=\textwidth]{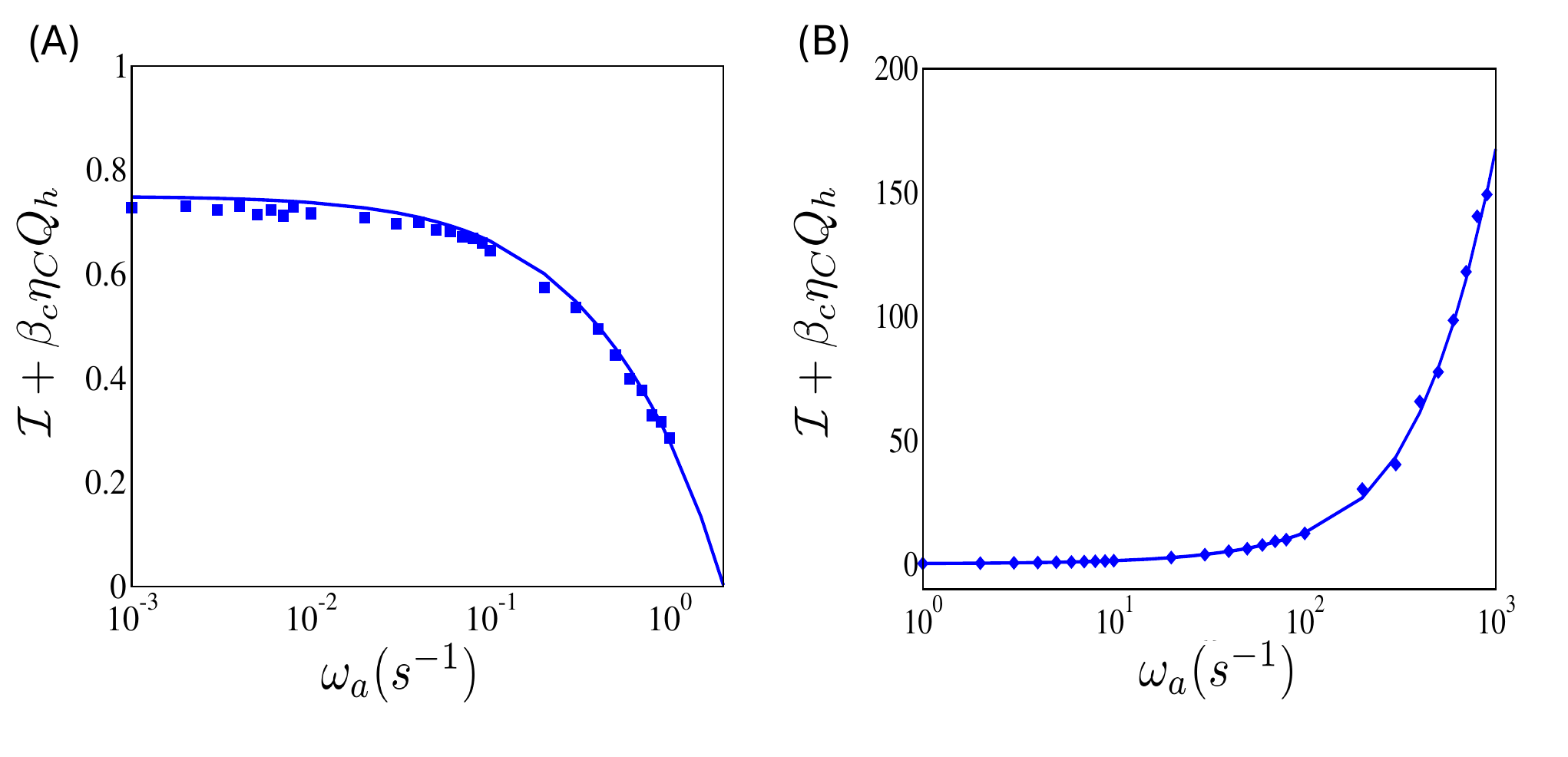}
\caption{Quasistatic thermodynamic quantity 
\(\mathcal{I}+\beta_c\eta_C Q_h\) for the two-stroke active heat engine as a function of the active shot rate \(\omega_a\).
We compare two cases where homogeneous and heterogeneous activity are present throughout the cycle. In both cases \(\omega_a\) is varied and all other parameters are kept fixed. (A) Homogeneous activity is present with a uniform active temperature ($T_{ah}=T_{ac}=144.9\,\mathrm{K}$). Other parameters are same as in Fig.~\ref{FIG4}. (B) Heterogeneous activity is present in terms of different active temperatures \(T_{ac}=14.5\,\mathrm{K}\) and  \(T_{ah}=58.0\,\mathrm{K}\). Other parameters are same as Fig. \ref{FIG5}.} 
\label{Fig:__plus_work_engine_regime_activity_overall}
\end{figure*}
We therefore examine the dependence of this quantity on the activity rate $\omega_a$ for different choices of the passive and active bath parameters.
Figure~\ref{Fig:__plus_work_engine_regime_activity_overall} shows $\mathcal{I}+\beta_c\eta_C Q_h $ as a function of $\omega_a$ for two representative cases. Panel~(A) corresponds to homogeneous activity, with ($\bar{\kappa}>1$), ($\overline{T}>1$), and ($\overline{T_a}=1$). Panel~(B) corresponds to heterogeneous activity, with ($\bar{\kappa}>1$), ($\overline{T}>1$), and ($\overline{T_a}>1$). In both cases, the plotted quantity remains positive over the activity range considered, in agreement with the condition required for the active-efficiency bound.
In the homogeneous case (A), the system operates as an engine only up to a finite value of the activity rate. Beyond this range, the engine condition is no longer satisfied and the system crosses into a different operating regime. Correspondingly, $\mathcal{I}+\beta_c\eta_C Q_h $ decreases as $\omega_a$ is increased, although it remains non-negative in the plotted range. In contrast, for heterogeneous activity (B), the system remains in the engine regime over the full range of $\omega_a$ considered. In this case, $\mathcal{I}+\beta_c\eta_C Q_h $ increases monotonically with activity, confirming that the active-efficiency bound remains satisfied.
\begin{figure*}[!htbp]
\centering
\includegraphics[width=\textwidth]{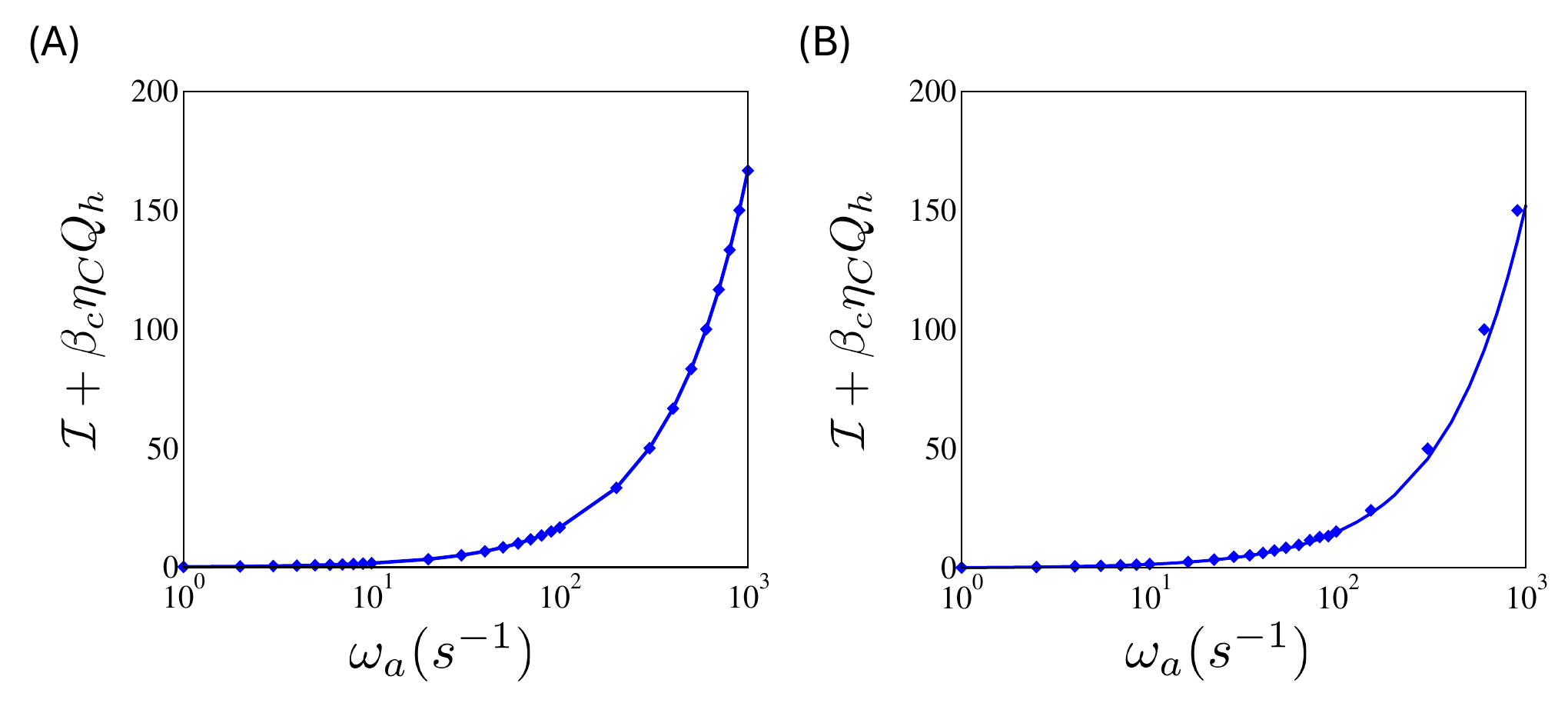}
\caption{Quasistatic thermodynamic quantity $(\mathcal{I}+\beta_c\eta_C Q_h)$ associated with the two-stroke active heat engine subject to thermal Gaussian white noise and active Poisson-shot noise baths. The engine operates in the quasistatic limit with a stepwise time-dependent PSN shot rate, corresponding to the third row of Table~\ref{TableRita}; see Sec.~\ref{sec:Vc} for further details. We show \(\mathcal{I}+\beta_c\eta_C Q_h\) as a function of the PSN shot rate \(\omega_a\), while all other parameters are kept fixed. (A) Case \(\kappa_1/\kappa_2=T_c/T_h\), with \(\kappa_2=0.3\,\mathrm{pN/nm}\).  (B) Case \(\kappa_1/\kappa_2<T_c/T_h\), with \(\kappa_2=0.32\,\mathrm{pN/nm}\).  The other parameters are $\tau_a=1.0\,\mathrm{s}$, $T_{ac}=14.5\,\mathrm{K}$, $T_{ah}=58.0\,\mathrm{K}$, $T_c=14.5\,\mathrm{K}$, $T_h=43.5\,\mathrm{K}$, $\gamma=0.2\,\mathrm{pN\,s/nm}$, $\kappa_1=0.1\,\mathrm{pN/nm}$,  cycle time $\tau=100\,\mathrm{s}$, number of simulations  $10^5$ and simulation time step $dt=0.1\,\mathrm{ms}$.}
\label{Fig:I_plus_effi_etaeqletac} 
\end{figure*}

We also check the same condition for the parameter regimes shown in Fig.~\ref{Fig:effi_etaeqletac}. The results are reported in Fig.~\ref{Fig:I_plus_effi_etaeqletac}. In both panels, $\mathcal{I}+\beta_c\eta_C Q_h $ remains positive throughout the range of $\omega_a$. This is consistent with the observation that the active efficiency remains bounded by the Carnot efficiency in this regime. Thus, the positivity of $\mathcal{I}+\beta_c\eta_C Q_h $ provides a direct confirmation that the active-efficiency bound is satisfied for the parameter ranges considered.

\begin{figure*}[!t]
\centering
\includegraphics[width=0.8\textwidth]{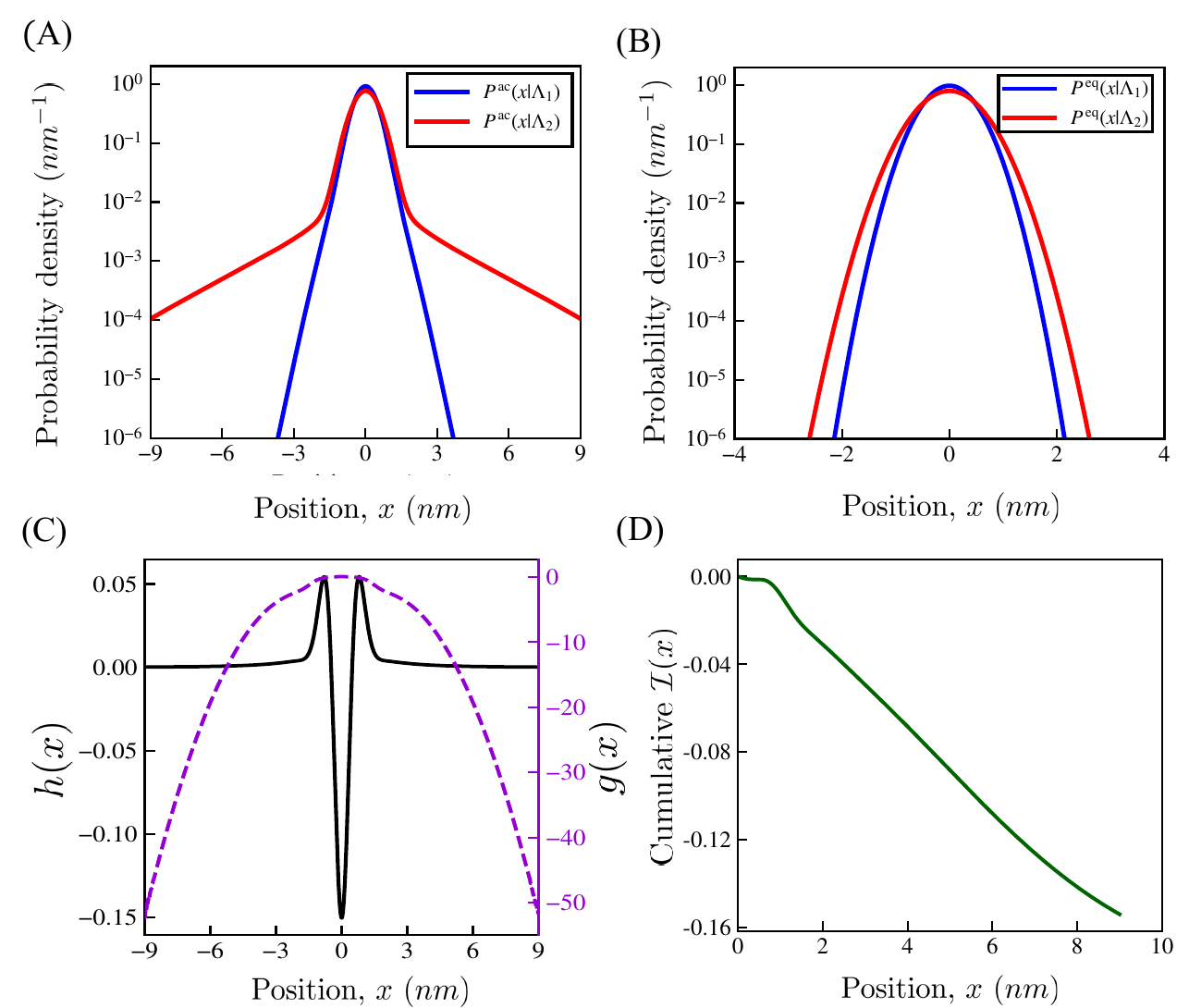}
\caption{Spatial contributions to the quasistatic divergence \(\mathcal I\) for the two-stroke active heat engine. (A) Accompanying distributions \(P^{\rm ac}(x|\Lambda_1)\) and \(P^{\rm ac}(x|\Lambda_2)\). (B) Corresponding equilibrium distributions \(P^{\rm eq}(x|\Lambda_1)\) and \(P^{\rm eq}(x|\Lambda_2)\). (C) Difference between the accompanying distributions, \(h(x)=P^{\rm ac}(x|\Lambda_2)-P^{\rm ac}(x|\Lambda_1)\), and the logarithmic factor \(g(x)=\log[P^{\rm ac}(x|\Lambda_2)/P^{\rm eq}(x|\Lambda_2)]-\log[P^{\rm ac}(x|\Lambda_1)/P^{\rm eq}(x|\Lambda_1)]\). (D) Cumulative spatial integral \(\mathcal I(x)=2\int_0^{x}h(x')g(x')\,dx'\), showing the contributions accumulated over the displayed spatial range. The distributions are calculated analytically using Eqs. (13) and (16). Parameters are \(\kappa_1=0.6\) pN/nm, \(\kappa_2=4.0\) pN/nm, \(T_c=T_{ac}=7.25\) K, \(T_h=72.5\) K, \(T_{ah}=362.3\) K, \(\gamma=0.2\) pN s/nm, \(\tau_a=1.0\) s, and \(\omega_a=0.5\) s\(^{-1}\).}
\label{FIGnew}
\end{figure*}

\subsubsection{Spatial contributions to the quasistatic divergence}
\label{app:spatial_divergence}
In Sec.~\ref{sec:Two_stroke_engine2}, we showed that the quasistatic divergence $\mathcal I$ accumulates only at the two switching events of the protocol. As illustrated in Fig.~\ref{fig:cumulative_I}, the two switches can contribute with opposite signs, and the final sign of $\mathcal I$ depends on which contribution dominates. To better understand the origin of these signs, we examine how the underlying probability distributions contribute to the quasistatic divergence. We start from Eq.~\eqref{I_general_psn}, which expresses $\mathcal I$ in terms of the accompanying active distributions and their corresponding equilibrium distributions. For convenience, we introduce the functions
\begin{equation}
\begin{aligned}
h(x)&=P^{\rm ac}(x|\Lambda_2)
     -P^{\rm ac}(x|\Lambda_1),~~
g(x)&=
\log\frac{P^{\rm ac}(x|\Lambda_2)}
{P^{\rm eq}(x|\Lambda_2)}
-
\log\frac{P^{\rm ac}(x|\Lambda_1)}
{P^{\rm eq}(x|\Lambda_1)}.
\end{aligned}
\label{eq:h_g}
\end{equation}
The quasistatic divergence can then be written as
\begin{equation}
\mathcal I=
\int_{-\infty}^{\infty}h(x)g(x)\,dx.
\label{eq:I_spatial}
\end{equation}
This expression allows us to examine how different regions of position space contribute to the sign of $\mathcal I$. Figure~\ref{FIGnew}.~(A) shows the accompanying active distributions for the two protocol states, while Fig.~\ref{FIGnew}.~(B) shows the corresponding equilibrium distributions. The functions $h(x)$ and $g(x)$, constructed from these distributions, are shown in Fig.~\ref{FIGnew}.~(C). Unless the two distributions are identical, $h(x)$ contains both positive and negative regions. The logarithmic factor $g(x)$ also has no general positivity constraint. For the
parameters considered here, both functions change sign at different positions, resulting in positive and negative contributions to the integrand $h(x)g(x)$. Therefore, the overall sign of $\mathcal I$ depends on the balance between these spatial contributions.

To examine how these contributions accumulate with position, we define the cumulative spatial integral
\begin{equation}
\mathcal I(x)=
2\int_0^x h(x')g(x')\,dx,
\label{eq:I_spatial_cumulative}
\end{equation} In the limit $x\to\infty$, this expression recovers the full quasistatic divergence. Figure~\ref{FIGnew}~(D) shows the cumulative integral as a function of $x$. The curve initially
decreases, followed by a small increase over an intermediate spatial region, before decreasing again. This nonmonotonic behavior reflects the competition between regions where $h(x)g(x)$
is positive and negative. The plotted range illustrates how these contributions accumulate, although the asymptotic value is not inferred from the finite spatial interval shown. This analysis complements the time-dependent picture presented in Fig.~\ref{fig:cumulative_I}. While the latter identifies the switching events responsible for changes in $\mathcal I$, the spatial decomposition shows how the differences between the accompanying active and equilibrium distributions contribute to its sign. In particular, the sign of $\mathcal I$ cannot be determined from either factor separately, but depends on their combined contributions over position space.

\section{Stirling protocol}
\label{app:stirling_protocol}
While the main text is devoted to piecewise-constant engine cycles, the analytical solution derived in Eqs.~\eqref{eqn:FokkerPlanck_solution}-\eqref{eqn:varphi_def2} holds for arbitrary time dependencies of the control parameters $\Lambda(t)=\{\kappa(t),T(t),T_a(t)\}$. We therefore consider here a Stirling-type engine with a smoothly varying trap stiffness $\kappa(t)$ in order to illustrate the broader applicability of the formalism.  The functional form of the driving associated with the stiffness is   
\begin{equation}
\kappa(t) = 
\begin{cases}
\displaystyle\frac{\kappa_1\kappa_2}{\displaystyle \kappa_2 + (\kappa_1 - \kappa_2) \frac{2t}{\tau}}, & 0 \leq t \leq \tau/2 \\\\
\displaystyle\frac{\kappa_1\kappa_2}{\displaystyle \kappa_1 + (\kappa_1 - \kappa_2) \left(1 - \frac{2t}{\tau} \right)}, & \tau/2 < t \leq \tau,
\end{cases}
\label{eq:TD_protocol}
\end{equation}
where \( \kappa_1 \) and \( \kappa_2 \) are respectively the minimum and maximum stiffness values, and \( \tau \) is the  cycle time. Furthermore, we take the  temperature $T(t)$ and PSN effective temperature $T_a(t)$  to be step-wise like in the two-stroke protocol given in Eq.~(\ref{maximum_efficiency_protocol}) of the Main Text.
\begin{figure}[t]
\begin{center}
\includegraphics[width=\textwidth]{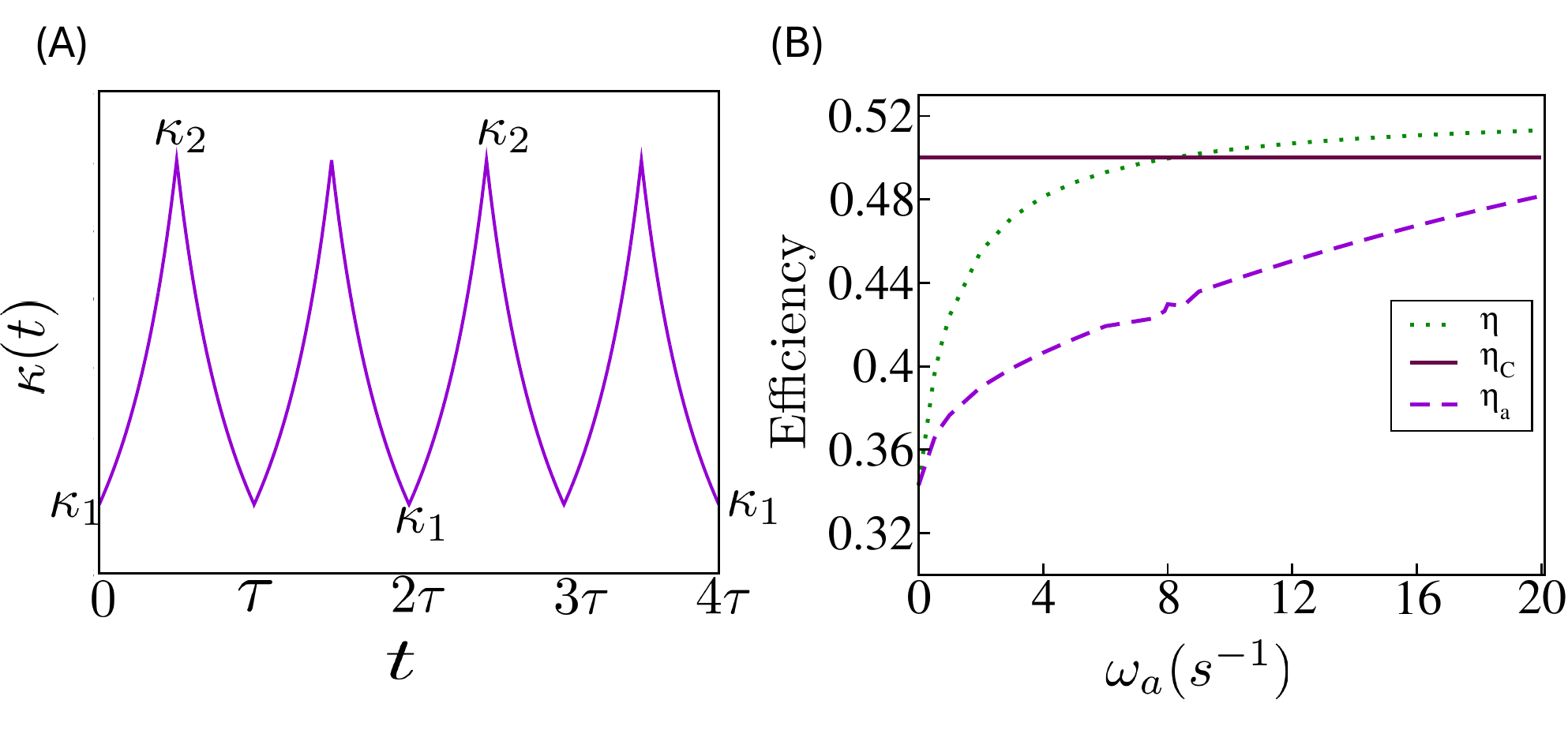}\hfill
\caption{Stirling-type active heat engine with smoothly varying trap stiffness. (A) Protocol for the time-dependent trap stiffness \(\kappa(t)\), which varies between \(\kappa_1\) and \(\kappa_2\) over a cycle of duration \(\tau\). (B) Traditional efficiency \(\eta=W/Q_h\) is shown with simulations as a green dotted line. The `active' efficiency \(\eta_a\) is shown with simulations in purple dashed line. The maroon line indicates Carnot's efficiency \(\eta_C=1-(T_c/T_h)\), defined in terms of the temperatures of the thermal baths. The parameters used are \(\tau_a=1.0\,\mathrm{s}\), \(T_{ac}=0.0\,\mathrm{K}\), \(T_{ah}=3.62\,\mathrm{K}\), \(T_c=1.45\,\mathrm{K}\), \(T_h=2.90\,\mathrm{K}\), \(\kappa_1=0.1\,\mathrm{pN/nm}\), \(\kappa_2=0.3\,\mathrm{pN/nm}\), \(\gamma=0.01\,\mathrm{pN\,s/nm}\), and cycle duration \(\tau=100\,\mathrm{s}\). The number of stochastic trajectories is \(10^5\), and the simulation time step is \(dt=0.1\,\mathrm{ms}\).}
\label{Fig.7}
\end{center}
\end{figure}

 We compute analytical expressions  for the variance and use it to calculate the key thermodynamic quantities associate with the Stirling cycle. The values of $\langle x^2(0)\rangle$ and $\langle x^2(\tau/2)\rangle$ can be calculated using the periodic condition and continuity properties of $\langle x^2(\tau/2^-)\rangle=\langle x^2(\tau/2^+)\rangle$ and $\langle x^2(0)\rangle=\langle x^2(\tau)\rangle$. We solve the Langevin equation Eq.~\eqref{eqn:Langevin_2} and define a parameter $a=\frac{\tau \kappa_1\kappa_2}{2\gamma(\kappa_1-\kappa_2)}$ for simplification of our calculation. The variance in the interval $0 \leq t \leq \tau/2$ reads,
\begin{align}
\langle x_{1}^{2}(t)\rangle
&=
\langle x^{2}(0)\rangle
\left[
\frac{\kappa_2+(\kappa_1-\kappa_2)\frac{2t}{\tau}}{\kappa_2}
\right]^{-2a}
+
\frac{\kappa_2\tau}
{(\kappa_1-\kappa_2)(2a+1)}
\left(
\frac{k_{B}T_{c}}{\gamma}
+
\frac{\omega_a \langle Y^{2}\rangle_{ac}}{2}
\right)
\nonumber\\
&\quad\times
\left[
\frac{\kappa_2+(\kappa_1-\kappa_2)\frac{2t}{\tau}}{\kappa_2}
-
\left(
\frac{\kappa_2+(\kappa_1-\kappa_2)\frac{2t}{\tau}}{\kappa_2}
\right)^{-2a}
\right].
\label{Eq.22}
\end{align}
The  variance in the second half of the cycle during $\tau/2 \leq t \leq \tau$ reads,

\begin{align}
\langle x_{2}^{2}(t)\rangle
&=
\langle x^{2}(\tau/2)\rangle
\left[
\frac{
\kappa_1+(\kappa_1-\kappa_2)\left(1-\frac{2t}{\tau}\right)
}{
\kappa_1
}
\right]^{2a}
+
\frac{
\kappa_1\tau
}{
(\kappa_1-\kappa_2)(2a-1)
}
\left(
\frac{k_{B}T_{h}}{\gamma}
+
\frac{\omega_a \langle Y^{2}\rangle_{ah}}{2}
\right)
\nonumber\\
&\quad \times
\left[
\frac{
\kappa_1+(\kappa_1-\kappa_2)\left(1-\frac{2t}{\tau}\right)
}{
\kappa_1
}
-
\left(
\frac{
\kappa_1+(\kappa_1-\kappa_2)\left(1-\frac{2t}{\tau}\right)
}{
\kappa_1
}
\right)^{2a}
\right].
\label{Eq.23}
\end{align}

 We also calculate the average total extracted work for an arbitrary cycle duration $\tau$ using Eq.~\eqref{eq:active_heat}. The total work is written as the sum of the contributions from the two halves of the cycle,
\begin{eqnarray}
    W=W_1+W_2.
    \label{eqn:td_work}
\end{eqnarray}
Where $W_1$ and $W_2$ denote the work performed during the first and second halves of the cycle, respectively. Both contributions contain passive thermal terms as well as active-noise terms. The active-noise contributions are determined by the second moments of the kick-amplitude distributions $\langle Y^2 \rangle_{ac}$ and $\langle Y^2 \rangle_{ah}$ whose expressions are given by Eq.~\eqref{eq:C1} and Eq.~\eqref{eq:C2}. 
\begin{align}
W_{1}
&=
\frac{
\langle x^2(0) \rangle \kappa_1 \kappa_2^{2 a+1}
\left(\kappa_1^{-2 a-1}-\kappa_2^{-2 a-1}\right)
}{
2 (-2 a-1)
}
+
\frac{
\kappa_1 \kappa_2 \tau
}{
2 (2 a+1)(\kappa_1-\kappa_2)
}
\left(
\frac{k_b T_c}{\gamma}
+
\frac{\omega_a \langle Y^{2}\rangle_{ac}}{2}
\right)
\left[
\ln \left(\frac{\kappa_1}{\kappa_2}\right)
-
\frac{
\left(\frac{\kappa_1}{\kappa_2}\right)^{-2 a-1}-1
}{
-2 a-1
}
\right].
\label{eq:w_td1}
\end{align}
In this expression, the term proportional to $\langle Y^{2}\rangle_{ac}$ is the active contribution associated with the active-kick ensemble characterized by $T_{ac}$.
We formulate the work during the second half of the cycle is given by,

\begin{align}
W_2
&=
\frac{
\kappa_2 \langle x^2(\tau/2)\rangle
\left[
\left(\frac{\kappa_2}{\kappa_1}\right)^{2 a-1}-1
\right]
}{
2 (2 a-1)
}
-
\frac{
\kappa_1 \kappa_2 \tau
}{
2 (1-2a)(\kappa_1-\kappa_2)
}
\left(
\frac{k_b T_h}{\gamma}
+
\frac{\omega_a \langle Y^{2}\rangle_{ah}}{2}
\right)
\left[
\ln\left(\frac{\kappa_2}{\kappa_1}\right)
-
\frac{
\left(\frac{\kappa_2}{\kappa_1}\right)^{2 a-1}-1
}{
2 a-1
}
\right].
\label{eq}
\end{align}
Here, the active contribution involves $\langle Y^{2}\rangle_{ah}$, since the second half of the cycle is associated with the active ensemble characterized by $T_{ah}$.
We calculate the expression of $Q_h$ using Eq.~\eqref{eq:active_heat} as,

\begin{align}
Q_{h}
&=
\frac{\kappa_1 \kappa_2\tau}{4\gamma^2(\kappa_1-\kappa_2)^2}
\Bigg[
\left(
2k_{B}T_h
+
\frac{\omega_a \langle Y^2\rangle_{ah}}{2}
\right)
\ln\left(\frac{\kappa_2}{\kappa_1}\right)
\left(
\frac{\kappa_1 \kappa_2 \tau}{2a-1}
+
\gamma (\kappa_2-\kappa_1)
\right)
\nonumber\\
&\qquad\qquad
+
\frac{
\tau
\left[
\kappa_2
-
\kappa_1
\left(\frac{\kappa_2}{\kappa_1}\right)^{2a}
\right]
}{
(1-2a)^2
}
\left[
2 \kappa_1 k_B T_h
+
\frac{\omega_a \langle Y^2\rangle_{ah}}{2}
-
(2a-1)\gamma
\langle x^2(\tau/2)\rangle
(\kappa_1-\kappa_2)
\right]
\Bigg].
\label{eq:td_heat1}
\end{align}
We calculate $Q_c$ using the Eq.~\eqref{eq:active_heat} and $\eta$ and $\eta_{a}$ using Eqs.~\eqref{eq:pseudo_efficiency} and~\eqref{eta_I}. Figure~\ref{Fig.7} summarizes, the main outcome of this time-dependent protocol. Figure~\ref{Fig.7}. (A)  shows the smoothly varying Stirling-type driving of the trap stiffness $\kappa(t)$ over successive cycles, interpolating between $\kappa_1$ and $\kappa_2$. Unlike for the piecewise-constant protocols discussed in the Main Text, this example provides a genuinely time-dependent realization of the engine, thereby illustrating the broader applicability of the analytical framework developed from the solution of the time-dependent Fokker-Planck equation. Figure~\ref{Fig.7}. (B) displays the corresponding efficiencies as functions of the activity $\omega_a$ when the active fluctuations are present only during the second half of the cycle. We find that the standard efficiency $\eta$ increases with activity and can exceed the Carnot value $\eta_C$ for moderate values of $\omega_a$, whereas the corrected active efficiency $\eta_a$ remains below $\eta_C$ throughout the parameter range explored in our simulations. This confirms, also for a smooth driving protocol, that the information-based correction plays a crucial role in restoring a thermodynamically meaningful bound in the presence of nonequilibrium active fluctuations.

\section{Numerical simulations}
\label{sec:code}
The expression given in Eq.~\eqref{eqn:ps_noneqsteady} was evaluated numerically. First, the quantity $J(q)$ defined in Eq.~\eqref{eqn:varphi_def} was computed on a finite grid of $q$ values. These discrete data points were then used to construct a smooth representation of $J(q)$ by means of cubic-spline interpolation. All integrations were carried out using an adaptive Gauss-Kronrod quadrature scheme. For the stochastic dynamics in Eq.~\eqref{eqn:Langevin_2}, stochastic trajectories were generated using a Euler-Maruyama integration scheme. The Gaussian white noise term $\xi(t)$ was implemented through increments of a Wiener process, while active kicks $\zeta_a(t)$ were introduced as random events occurring at Poisson-distributed times with rate $\omega_a$. The numerical probability density functions were estimated from the simulated trajectories. After discarding the initial transient regime, the sampled positions were grouped into bins, and normalized histograms were constructed to obtain the corresponding PDFs and the quasistatic divergence~$\mathcal{I}$, see Eq. ~\eqref{I_general_psn}.

\end{document}